\documentclass[11pt]{article}

\usepackage{slashed,cite}
\usepackage{graphicx, color} 
\usepackage{booktabs}
\usepackage{cleveref,url}

\pdfoutput=1
\usepackage{color}
\usepackage{amsmath,amssymb,hyperref,bm,bbold}
\usepackage{slashed,relsize,graphicx}
\def\beq{\begin{equation}\displaystyle}
\def\eeq{\end{equation}}
\def\bea{\begin{eqnarray}\displaystyle}
\def\eea{\end{eqnarray}}
\newcommand{\MET}{E\llap{/\kern1.5pt}_T}

\def\o{\vec{O}}
\def\s{\vec{s}}

\def\n{{\vec{\nu}}}

\def\e{\mathrm{e}}
\def\D{\vec{\Delta}}
\def\S{\textrm{\sc{S}}}
\def\B{\textrm{\sc{B}}}
\def\M{\textrm{\sc{M}}}

\def\Id{\mathbb{1}}
\def\mdm{m_{\textrm{\tiny{DM}}}}

\title{
\vspace{-2cm}
\begin{flushright}
\begin{flushright} \small{CERN-TH-2017-074}\end{flushright}
\end{flushright}
\vspace{3cm}
\bf \LARGE
Setting limits on Effective Field Theories:\\ the case of Dark Matter
\vspace{.2cm}}
\date{}
\author{
{\large Federico Pobbe$^{a}$, Andrea Wulzer$^{a,b,c}$, Marco Zanetti$^{a}$}\\
[5mm]
\normalsize\itshape $^a$ Dipartimento di Fisica e Astronomia, Universit\`a di Padova and\\ \normalsize\itshape INFN, Sezione di Padova, via Marzolo 8, I-35131 Padova, Italy\\
\normalsize\itshape $^b$ Theoretical Physics Department, CERN, Geneva, Switzerland\\
\normalsize\itshape $^c$ Theoretical Particle Physics Laboratory, Institute of Physics, \\ \normalsize\itshape EPFL, Lausanne, Switzerland\\
}

\date{}
\begin{document}
\baselineskip=13pt
\maketitle
\begin{abstract}
The usage of Effective Field Theories (EFT) for LHC new physics searches is receiving increasing attention. It is thus important to clarify all the aspects related with the applicability of the EFT formalism in the LHC environment, where the large available energy can produce reactions that overcome the maximal range of validity, i.e. the cutoff, of the theory. We show that this does not forbid to set rigorous limits on the EFT parameter space through a modified version of the ordinary binned likelihood hypothesis test, which we design and validate. Our limit-setting strategy can be carried on in its full-fledged form by the LHC experimental collaborations, or performed externally to the collaborations, through the Simplified Likelihood approach, by relying on certain approximations. We apply it to the recent CMS mono-jet analysis and derive limits on a Dark Matter (DM) EFT model. DM is selected as a case study because the limited reach on the DM production EFT Wilson coefficient and the structure of the theory suggests that the cutoff might be dangerously low, well within the LHC reach. However our strategy can also be applied, if needed, to EFT's parametrising the indirect effects of heavy new physics in the Electroweak and Higgs sectors.
\end{abstract}


\newpage
\section{Introduction}

The preliminary LHC results made clear that new physics does not assume the ``vanilla'' form we had imagined for it, leaving us uncertain on how to proceed in the search for new phenomena. The problem is that no single explicit new physics model, through which complete signal predictions could be made, singles out at present as compelling or particularly motivated. Not having to our disposal a sharp signal hypothesis to be compared with the data is obviously a complication. The problem has been addressed, and solved in most cases, by developing comprehensive phenomenological descriptions aimed at parametrising a wide set of new physics models or scenarios in terms of few free parameters. The transition from the constrained MSSM to phenomenological parametrisations of the signal topologies broadly expected in SUSY, advocated in ref.~\cite{ArkaniHamed:2007fw}, is the paradigmatic example of this approach. 

Phenomenological models are useful, but they must be employed with care. It must be kept in mind that they own their simplicity and generality to the fact that they are designed to describe only a specific set of physical processes. Their few free parameters indeed correspond to those combinations of the underlying theory parameters that control the processes we decided to focus on. If used outside their domain of validity, their predictions might be incompatible with the new physics scenario they aim at describing, or they correspond to a very specific model and loose their generality. For instance, it is obvious that one should not try to study Dark Matter (DM) in SUSY through a model designed to describe the gluino pair-production topology. Or, one should not use a model for vector resonances production \cite{Pappadopulo:2014qza} to describe the shape of the signal outside the resonant peak. Making a proper usage of phenomenological models is simple in most cases: it suffices to select for the analysis only those events that are within the domain of validity of the model. However this might not be possible, for instance because the events to which the model applies are not experimentally distinguishable from the others. When this occurs, the situation is more complicated and a special treatment is needed. Addressing this issue for DM searches based on Effective Field Theories (EFT) is the purpose of the present paper. The strategy we will develop can be also applied, if needed, to other EFT-based analyses. 

The formalism of low-energy EFT's allows us to produce phenomenological models in a rigorous and systematic way. EFT's provide a complete description of the low-energy dynamics of the lightest particles of a given microscopic (UV) theory, under the sole assumption that the rest of the spectrum sits at an energy scale $M_{\textrm{cut}}$ which is much above the light particles' mass. The validity domain of EFT-based phenomenological models is thus not restricted to a limited set of topologies, but rather to a limited range of energy. All reactions with a center of mass energy $E_{\rm{cm}}$ below $M_{\textrm{cut}}$, called the EFT cutoff, are accurately described by the EFT in terms of a few leading EFT interaction operators. The coefficients of these operators, called Wilson coefficients, are free parameters of the EFT. Any UV theory with appropriate low-energy spectrum corresponds to a given value of the Wilson coefficients. By varying the latter ones can thus effectively explore all the UV models with the light particles (and symmetries, which restrict the allowed interaction operators) content we assumed in the construction of the EFT. One can of course also reverse the logic and, starting from experimental results (e.g., from exclusion limits) given in terms of the Wilson coefficients, translate them into constraints on any specific model, in which the Wilson coefficients can be calculated. All this however only works if the EFT is properly employed, i.e. if the comparison of the theory with data relies in no way on the EFT predictions for reactions with $E_{\rm{cm}}$ at or above the cutoff $M_{\textrm{cut}}$. As $E_{\rm{cm}}$ approaches $M_{\textrm{cut}}$, indeed, the EFT description obtained with the leading operators becomes less and less accurate and more and more sub-leading interactions should be added. At $E_{\rm{cm}}\geq M_{\textrm{cut}}$, infinitely many operators become relevant and the leading-order EFT prediction looses any resemblance to the one of the original theory. In the high energy region the EFT predictions should be replaced with those of the underlying UV theory, which however is precisely what we want to be agnostic about.

EFT's have been employed for new physics searches in a multitude of different contexts. Applications range from flavour and electroweak (EW) precision physics to DM direct and indirect searches and, more recently, to LHC studies of DM \cite{Goodman:1984dc,Beltran:2008xg,Cao:2009uw,Beltran:2010ww,Goodman:2010yf,Bai:2010hh,Fan:2010gt,Goodman:2010ku} and of the EW-plus-Higgs sector \cite{Pomarol:2013zra,Buchalla:2013rka,Wells:2014pga,Batell:2012ca,Alonso:2012px,Contino:2013kra,Ellis:2014jta}. The limited range of validity of the EFT is never an issue for non-LHC applications, because the experimental conditions automatically forbid $E_{\rm{cm}}$ to exceed $M_{\rm{cut}}$. For instance in flavour measurements, $E_{\rm{cm}}<$~few~GeV, while the new physics scale one probes by the EFT (i.e., $M_{\rm{cut}}$) is easily above or much above the TeV. A lot of energy is instead available at the LHC and the EFT validity problem must be carefully addressed. This has been studied in ref.s~\cite{Goodman:2011jq,Shoemaker:2011vi,Profumo:2013hqa,Busoni:2013lha,Busoni:2014sya,Busoni:2014haa,Racco:2015dxa,Matsumoto:2014rxa,Matsumoto:2016hbs} in the context of DM and in ref.s~\cite{Contino:2016jqw,Farina:2016rws} for EW-plus-Higgs sector EFT's. The cases discussed in ref.~\cite{Contino:2016jqw}, and in ref.~\cite{Farina:2016rws} for the neutral Drell--Yan analysis, are very easy to deal with, as $E_{\rm{cm}}$ can be measured experimentally. By an $E_{\rm{cm}}<M_{\textrm{cut}}$ selection cut one can thus restrict the data to the region where the EFT predictions are reliable. With this cut in place, data can be compared with the EFT prediction with the standard statistical tools (see e.g.~\cite{Cowan:1998ji,Agashe:2014kda,Rolke:2004mj,Cowan:2010js}) producing measurements of the Wilson coefficients or, more often, exclusion limits.\footnote{Notice that the limits depend on $M_{\textrm{cut}}$, which is a free parameter to be scanned over. Limits are thus effectively set in the enlarged parameter space defined by the Wilson coefficients plus the EFT cutoff $M_{\textrm{cut}}$. This is perfectly correct because the cutoff is indeed one of the parameters of the EFT, as extensively discussed in \cite{Racco:2015dxa}.} Obviously $E_{\rm{cm}}$ cannot be measured in DM searches because the DM particles escape from the detector. Still we will show that a fully rigorous limit-setting strategy can be set up and used to interprete LHC ``mono-$X$'' DM searches in the EFT context. Our approach is instead not suited to characterise a discovery. We will comment on this point in the Conclusions.

Mono-$X$ DM searches consist in measuring the detector-level Missing Transverse Energy ($\MET$) distribution in events characterised by the presence of one visible object $X$. Concretely, $X$ can be a jet \cite{Aaboud:2016tnv,Sirunyan:2017hci}, a photon \cite{Aaboud:2016uro}, a vector or Higgs boson \cite{Aaboud:2016obm,Sirunyan:2017hnk,Sirunyan:2017onm} or even a top pair \cite{Khachatryan:2015nua}. The production of DM contributes to the $\MET$ distribution, or better to its binned version measured in $N$ bins, by an amount
\beq\label{dist}
n_i^{\textrm{\tiny{DM}}}=\hspace{-4pt}\int\limits_{\textrm{bin}_i}\hspace{-3pt}d\MET\frac{dn_i^{\textrm{\tiny{DM}}}}{d\MET}=
\hspace{-4pt}\int\limits_{\textrm{bin}_i}\hspace{-3pt}d\MET
\hspace{-4pt}\int^{\tiny\sqrt{S}}\hspace{-8pt}dE_{\rm{cm}}\frac{d^2\hspace{-1pt}n_i^{\textrm{\tiny{DM}}}}{dE_{\rm{cm}}d\MET}\,,
\eeq
where $d^2\hspace{-1pt}n_i^{\textrm{\tiny{DM}}}$ is the detector-level DM$+X$ production cross-section times the integrated luminosity, doubly differential in $\MET$ and in the center of mass energy.\footnote{An unambiguous definition of $E_{\rm{cm}}$ is postponed to Section~\ref{DMEFTLIM}.} The equation has been written in a format that outlines how the DM production signal in each $\MET$ bin receives contributions from all the $E_{\rm{cm}}$ energy scales up to the total collider energy $\sqrt{S}=13\,$TeV at the LHC run-$2$. Notice instead that a lower bound (depending on $\MET$) can be set on the $E_{\rm{cm}}$ integration domain. This is not relevant for the discussion and thus it has not been reported in the equation. 

The problem with the EFT is that it does not allow us to compute eq.~(\ref{dist}) fully, but only part of it. Namely, splitting the $E_{\rm{cm}}$ integral in the regions below and above the cutoff, we have
\beq\label{dist1}
n_i^{\textrm{\tiny{DM}}}
=
\hspace{-4pt}\int\limits_{\textrm{bin}_i}\hspace{-3pt}d\MET
\hspace{-4pt}\int^{\tiny{M_{\rm{cut}}}}\hspace{-8pt}dE_{\rm{cm}}\frac{d^2\hspace{-1pt}n_i^{\textrm{\tiny{DM}}}}{dE_{\rm{cm}}d\MET}+
\hspace{-4pt}\int\limits_{\textrm{bin}_i}\hspace{-3pt}d\MET
\hspace{-4pt}\int_{\tiny{M_{\rm{cut}}}}^{\tiny\sqrt{S}}\hspace{-8pt}dE_{\rm{cm}}\frac{d^2\hspace{-1pt}n_i^{\textrm{\tiny{DM}}}}{dE_{\rm{cm}}d\MET}
\equiv n_i^{\textrm{\tiny{EFT}}} +\Delta n_i\,.
\eeq
The first term accounts from DM production occurring at $E_{{\rm{cm}}}<{M_{\rm{cut}}}$, i.e. within the domain of validity of the EFT. We can thus compute it by using the EFT, as a function of its free parameters (including the cutoff), obtaining the prediction $n_i^{\textrm{\tiny{EFT}}}$. The second contribution, $\Delta n_i$, is instead not known from the viewpoint of the EFT because it accounts for ``hard'' DM production, originating from energy scales where the EFT description no longer applies. All what we know for sure is that it is necessarily positive, $\Delta n_i\geq0\;\forall i$. This being so is obvious from the definition, and physically due to the fact that $n_i^{\textrm{\tiny{EFT}}}$ and $\Delta n_i$ emerge from two kinematically different (and thus quantum-mechanically distinguishable) processes. As such they cannot interfere, producing two positive contributions that add up to each other. In what follows, we will denote $n_i^{\textrm{\tiny{EFT}}}$ as ``the signal'', in spite of the fact that it is only one of the two contributions to the total DM production rate. We will refer to the second contribution, $\Delta n_i$, as ``unknown additional signal''.

The statistical problem of setting limits in the presence of an unknown additional signal has never been seriously addressed in the literature. A simple but unsatisfactory solution (adopted for instance in ref.~\cite{Racco:2015dxa}) is to turn the experimental analysis into a cut-and-count experiment, i.e. to measure the distribution in a single $\MET$ bin. Together with the estimate of the SM background, this single measurement produces an upper limit, call it ${\overline{n}}^{\textrm{\tiny{DM}}}_{\textrm{\tiny{exc.}}}$, on the total DM production signal (or, equivalently, on the DM production cross-section) one can tolerate in the search region with a given (typically, $95\%$) Confidence Level (CL). The limit is fully model-independent and it is readily interpreted in the EFT by writing
\beq\displaystyle\label{singlebin}
{\overline{n}}^{\textrm{\tiny{EFT}}} \leq {\overline{n}}^{\textrm{\tiny{DM}}}\leq {\overline{n}}^{\textrm{\tiny{DM}}}_{\textrm{\tiny{exc.}}}\,.
\eeq
This method is rigorous, but largely unsatisfactory. First because the choice of the search region needs to be optimised not only for each given EFT model, but also for each point in the EFT parameter space. This was shown in ref.~\cite{Racco:2015dxa}, where the $4$ search regions considered by ATLAS in ref.~\cite{ATLAS:2012zim} were found to have different limit-setting performances in different regions of the EFT parameter space. Moreover, the cut-and-count approach is never fully optimal as it does not make use of the whole available experimental information. For that, one would need to employ the entire measured distribution, but this is not straightforward in the presence of additional signal. The standard statistical methods to compare a binned histogram with the data crucially rely on the ability of the model to provide a definite prediction for (at least) the \emph{shape} of the distribution. The EFT model instead gives us only a \emph{lower limit} on the distribution and tells us nothing about the shape, which is sensitive to the one of the additional signal component. A slight variation of the standard hypothesis test methodology is needed in order to deal with this situation. 

The rest of the paper is organised as follows. In sect.~\ref{hyp} we discuss several strategies to address the problem of limit-setting in the presence of additional signal. We start from a complete approach, to be adopted by the LHC experimental collaboration that have access to all the details of the analysis, and simplify it down to a level that allows it to be performed outside the collaborations, relying on the minimal amount of experimental information. The usage of simplified likelihoods \cite{Collaboration:2242860} (see also \cite{Fichet:2016gvx}) emerges as a valid strategy, but alternatives could also be considered. In sect.~\ref{DMEFTLIM} we return to DM EFT and set limits on a specific $d=6$ EFT operator by re-interpreting the CMS mono-jet analysis in ref.~\cite{Sirunyan:2017hci}, for which the simplified likelihood is available. On top of serving as an illustration of the limit-setting method, this example will give us the opportunity to introduce and to validate the operative definition of the center of mass energy $E_{{\rm{cm}}}$ to be employed for the $E_{{\rm{cm}}}<{M_{\rm{cut}}}$ restriction in the EFT signal simulation. In sect.~\ref{conc} we report our conclusions and comment on the applicability of our study outside the domain of DM searches. A validation of our limit-setting strategy and some algebraic derivations are reported in appendices~\ref{toy} and \ref{appB}, respectively.

\section{Hypothesis test with unknown additional signal}\label{hyp}

\subsection{The Problem}\label{thp}

Consider measuring $N$ observables $\o=\{O_i\}$, $i=1,\ldots N$, whose joint p.d.f. (the probability model) depends on $N$ ``known'' quantities $\S_i$ (the signal), plus $N$ unknown additional signal components $\Delta_i\geq0$. Namely, the likelihood function is
\beq\label{lik0}
L=L(\vec\S,\D ;\o\,)={\mathcal{P}}(\o\,|\vec\S+\D)\,.
\eeq
In what follows we will mostly be interested in interpreting $\o$ as event countings in $N$ bins, $\vec\S$ as the expected yields as predicted by the EFT and $\D$ as the additional signal yield from reactions above the EFT cutoff, in direct correspondence with eq.~(\ref{dist1}). However $\o$ might also represent parton-level differential cross-section measurements, in which case $\vec\S$ and $\D$ are the cross-section predictions from low and high energy processes, respectively. Other parameters $\n$, among which the backgrounds, also influence the probability model. These are routinely treated as nuisance parameters (see e.g., \cite{Rolke:2004mj,Cowan:2010js}) and constrained by means of auxiliary measurements. Nuisance parameters play no role for the conceptual point we are making here, we thus momentarily ignore their presence.

In order to construct an hypothesis test one defines in some way (typically out of the likelihood function) a test statistic 
\beq
t(\vec\S,\D;\o\,)\,,
\eeq
which is such that a large value of $t$ signals tension of the model with observations. The test statistic is itself a random variable, because of its dependence on $\o$. Its distribution follows from the probability model for $\o$ and therefore it depends, a priori, on $\vec\S$ and $\D$ in a highly non trivial manner. Namely
\beq
t\sim f(t\,|\vec\S,\D)\,.
\eeq
The probability for the model to produce an experimental result which is equally or more incompatible than the one observed is quantified by the $p$-value
\beq
\label{pv0}
p(\vec\S,\D;\o\,)=\int_{t(\vec\S,\D;\o\,)}^\infty \hspace{-5pt}dt'\, f(t'|\vec\S,\D)\,.
\eeq
The model is said to be excluded at a given CL $\alpha$ if $p<1-\alpha$, it is not excluded otherwise.

The hypothesis test defined by eq.~(\ref{pv0}) is of course not what we want. It is a test for the probability model of $\o$, which requires both $\vec\S$ and $\D$ to be  specified. Our goal is instead to test a physics model that only predicts $\vec\S$ and leaves complete freedom on what the additional signal $\D$ could be. We thus want our test to be such that the model is excluded only if its prediction are incompatible with the data for any value of $\D$, provided of course $\Delta_i\geq0\;\forall i$.\footnote{Each given UV-completion of the EFT model corresponds to a given value of $\D$. Asking for the EFT to be excluded for any $\D$ thus ensures that all its possible UV-completions are excluded.} In practice, we want the model to be excluded \emph{only if} it fails the ``ordinary'' statistical test (i.e., $p<1-\alpha$ in eq.~(\ref{pv0})) \emph{for any} possible choice of the $\Delta_i$'s. We would also like our test to exclude as much as possible, namely we want the model \emph{not} to be excluded \emph{only if} there exists \emph{at least one} point in the $\D$-space that would pass the ordinary test. Both these requirements are fulfilled by defining the $p$-value as
\beq\label{pvid}
p_{\textrm{max}}(\vec\S;\o\,)=\underset{\Delta_i\geq0}{\textrm{\large{sup}}}\bigg\{p(\vec\S,\D;\o\,)\bigg\}
=\underset{\Delta_i\geq0}{\textrm{\large{sup}}}\bigg\{
\int_{t(\vec\S,\D;\o\,)}^\infty \hspace{-5pt}dt'\, f(t'|\vec\S,\D)
\bigg\}
\,,
\eeq
and use it to set a limit in the $\vec\S$-space depending on whether $p_{\textrm{max}}\lessgtr1-\alpha$.

Eq.~(\ref{pvid}) is in principle all what we need to set limits in the presence of unknown additional signal, however it is not applicable in concrete. The main obstruction is the need of determining the p.d.f. of the test statistic, which in general cannot be computed analytically and must be obtained numerically by running a toy Monte Carlo. The p.d.f. depends on $\D$, therefore one toy Monte Carlo would be needed at each point in the $\D$-space, resulting in a too demanding numerical procedure. Fortunately this problem is easy to address because the standard test statistic variables employed in LHC analyses are defined in such a way that their p.d.f.s assume a specific form (typically, a $\chi^2$ distribution) under suitable conditions, namely in the so-called Asymptotic Limit (AL). The AL approximation relies on the assumption that a large set of data is employed in the analysis. Further qualifications and checks of its validity are reported in the following section and in appendix~\ref{toy}. It suffices here to notice that in the AL the p.d.f. of $t$ becomes independent of $\D$ (and of $\vec\S$) so that eq.~(\ref{pvid}) becomes
\beq\label{pvAL}
p_{\textrm{max}}\overset{{\rm{AL}}}{=}
\underset{\Delta_i\geq0}{\textrm{\large{sup}}}\bigg\{
\int_{t(\vec\S+\D;\o\,)}^\infty \hspace{-5pt}dt'\, f(t')\bigg\}=\int_{t_{\rm{min}}}^\infty \hspace{-5pt}dt'\, f(t')\,,\;{\rm{where}}\;\;
t_{\textrm{min}}(\vec\S;\o\,)=\underset{\Delta_i\geq0}{\textrm{\large{inf}}}\{t(\vec\S,\D;\o\,)\}\,.
\eeq
The problem thus reduces to the one of minimising $t$ in the $\D$ space. We will show in the next section how to treat it by a combination of analytical and numerical techniques.

Before concluding this section it is worth noticing that it would be legitimate to regard the $\D$'s as additional nuisance parameters. Exactly like the $\n$'s, indeed, they are parameters whose values we are not interested in testing that however enter in the probability model for the observables. We decided not to adopt this interpretation and we introduced the terminology of ``unknown additional signal'' because ordinary nuisance parameters are constrained by auxiliary measurements (or by theoretical considerations, for theory-driven errors) while there is no way to constrain the $\D$'s.

\subsection{Possible solutions}

We discuss now how to deal with eq.~(\ref{pvid}) (actually with its much simpler version (\ref{pvAL})) in concrete. We begin with the full-fledged treatment, to be carried on directly by the experimental collaborations starting from the actual data and from the detailed knowledge of the nuisance parameters, and we progressively simplify it.

\subsubsection{The full story}\label{full}

In an LHC experimental analysis, the observations $\o$ are event countings in $N$ disjoint bins. Their p.d.f. is thus the product of $N$ independent Poisson distributions and it is fully specified by the mean values $\vec\M=\{\M_i\}$. The $\vec\M$'s consist of three terms
\beq\label{expnus}
\vec\M(\s,\D,\n)=\vec\S(\s,\n)+\vec\B(\n)+\D\,,
\eeq
where $\vec\S$ is the signal, $\vec\B$ is the background, $\D$ is the additional signal and $\n=\{\nu^{\hat{a}}\}$ ($\hat{a}=1,\ldots,\kappa$) represents the nuisance parameters. The signal is the component of $\vec\M$ that emerges from the new physics model we aim at testing, i.e. from the EFT with the $E_{{\rm{cm}}}<{M_{\rm{cut}}}$ restriction as in eq.~(\ref{dist1}).\footnote{The $E_{{\rm{cm}}}<{M_{\rm{cut}}}$ restriction must be applied to the complete physical signal event yield calculation. Specifically this means that the restricted prediction must include all the SM diagrams, if any, that interfere with the EFT vertex. More details in section~\ref{inter}.} It depends of course on nuisance parameters such as the uncertainties on the luminosity, on the efficiency for trigger and final state reconstruction, etc. Notice that the signal $\vec\S$ is taken here to depend on $N$ parameters $\s=\{s_i\}$, one for each histogram bin. Those are meant to be $N$ quantities of purely theoretical nature, calculable within our model with no reference to experimental effects that would induce a dependence on the nuisance parameters. For definiteness, we will think to $\s$ as the parton-level EFT cross-section in each bin, however we will never need to specify its exact definition because $\vec\S$ will be directly obtained from simulations performed within the model of interest, with no need of computing it as a function of $\s$ as an intermediate step. Notice that our approach is different from the standard one \cite{Agashe:2014kda,Rolke:2004mj,Cowan:2010js}. In the latter, one takes into account from the very beginning that $\vec\S$ is actually function of a smaller set of parameters, namely of the free parameters of the BSM model under consideration. One actually goes even further than that, and treats all the BSM parameters as fixed, aside from the total cross-section normalisation $\mu$ (the so-called signal-strength). What we are doing here is to first construct an hypothesis test for a generic model in which the cross-section in each bin is a free parameter, and next restrict it to the model of interest in which $\s$ (and in turn $\vec\S$) is function of $\mu$ only. The concrete implications of this different approach are described below and the reason why we did not adopt the standard strategy is explained in sect.~\ref{whynot}.

The other terms in eq.~(\ref{expnus}) are easier to understand. $\vec\B$ is the ``ideally reducible'' SM background (namely, the one that does not interfere with the signal, see section~\ref{inter}), some of whose components are obtained from simulations while others are measured in control regions. The background is affected by nuisance parameters which are not necessarily the same ones appearing in the signal prediction. For instance, the background components estimated from data are themselves nuisance parameters, which obviously do not affect the signal prediction. Our notation, in which all the nuisance are collected in a single vector $\n$, is general enough to account for all possibilities. The last term in eq.~(\ref{expnus}) is the additional signal component $\D$ and it is independent of the nuisance. One might find this confusing if thinking to the additional signal as emerging from an additional term in the parton-level cross-section. A detector simulation would be needed to relate this additional cross-section to the additional signal and a dependence on the nuisance would be introduced. However the additional cross-section is completely arbitrary (but positive) and therefore it produces a completely arbitrary (but positive, $\Delta_i\geq0$) additional signal. We can thus just take the latter, instead of the additional cross-section, to specify our probability model.

The complete likelihood function reads
\beq\label{likfull}
{{L}}[\vec\M(\s,\D,\n),\n;{\o}\,]=\prod\limits_{i=1}^{N}\frac{\M_i^{O_i}}{O_i!}\e^{-\M_i}\cdot{{L}}_\n\,,
\eeq
where ${{L}}_\n$ is the likelihood for the nuisance parameters obtained by a set of auxiliary measurements, such as those performed in control region or in detector calibration studies. Auxiliary measurements being by assumption insensitive to the putative signal, ${{L}}_\n$ is exclusively a function of $\n$. Out of the complete likelihood in eq.~(\ref{likfull}), the test statistic is defined as the Profile Log Likelihood Ratio 
\beq\label{PLR1}
\displaystyle
t(\s,\D;\o\,)=-2\log\bigg[
{\underset{\n}{\textrm{\large{sup}}}\big\{L(\vec\M,\n;\o\,)\big\}}\bigg/{\underset{\n,\s}{\textrm{\large{sup}}}\big\{L(\vec\M,\n;\o\,)\big\}}
\bigg]\,.
\eeq
In the numerator, the likelihood is maximised with respect to the nuisance parameters only, while keeping $\s$ fixed. In the denominator, one maximises with respect to both $\s$ and $\n$, leading to the absolute maximum of the likelihood in the entire parameter space of the $\s$'s and of the $\n$'s. No maximisation is instead performed over the $\D$'s that are treated, at this stage, as fixed constants. Our first goal is indeed to define a test for each individual $\D$ hypothesis, to be eventually turned into a $\D$-independent test with the strategy outlined in the previous section.

The test statistic in eq.~(\ref{PLR1}) perfectly complies with the general definition of Profile Log Likelihood Ratio given in \cite{Wilks:1938dza,Wald,Rolke:2004mj,Agashe:2014kda}, however it differs from the standard definition employed in LHC analyses \cite{Rolke:2004mj,Cowan:2010js}. In the standard case, as previously explained, the signal is restricted from the very beginning to be the one predicted by the model at hand and the signal-strength $\mu$ is taken as the only ``parameter of interest''. This means that the supremum in the denominator of the likelihood ratio formula is taken over $\mu$, rather than over the $N$ parameters $s_i$ as in our case. This difference has two implications. The first is that the denominator can be computed analytically in our case. Given that we now have $N$ free parameters to maximise over, we can use them to set $\M_i=O_i$ in each bin, reaching in this way the absolute maximum of the Poisson likelihood.\footnote{This might require setting some of the $s_i$'s to a negative value. Doing so means that we are constructing a test for an extremely general model in which the signal cross-sections $s_i$'s can even be negative, possibly due to some destructive interference. After constructing the test for such a general model we will apply it to the EFT, in which of course the $s_i$'s are positive.} At the same time one of course sets $\n$ to whatever value (call it $\n_0$) maximises $L_\n$. We can thus rewrite eq.~(\ref{PLR1}) as
\beq\label{PLR2}
\displaystyle
t=\hspace{-1pt}-2\log\hspace{-2pt}\bigg[
{\underset{\n}{\textrm{\large{sup}}}\{L(\vec\M,\n;\o\,)]\}}/L(\o,\n_0;\o\,)\hspace{-1pt}
\bigg]\hspace{-2pt}
=\underset{\n}{\textrm{\large{inf}}}\left\{\hspace{-1pt}2\hspace{-1pt}\sum\limits_{i=1}^N\left(\hspace{-1pt}\M_i\hspace{-1pt}-\hspace{-1pt}O_i\hspace{-1pt}-\hspace{-1pt}O_i\log\frac{\M_i}{O_i}\hspace{-1pt}\right)\hspace{-1pt}
-\hspace{-1pt}2\log\frac{L_\n}{L_{\n_0}}
\right\}.
\eeq
This formula is the generalisation, including nuisance, of the maximum likelihood goodness-of-fit test for Poisson countings (see e.g. \cite{Cowan:1998ji}). 

The second difference with the ordinary test is that $t$ is now distributed, in the AL, as a $\chi^2$ with $N$ degrees of freedom, unlike the ordinary ``signal-strength-based'' test statistic that is distributed as a $\chi^2$ with $1$ degree of freedom independently of the number of bins. This result straightforwardly follows from the Wilks' theorem \cite{Wilks:1938dza,Wald}, according to which the Profile Log Likelihood Ratio follows a $\chi^2$ distribution with a number of degrees of freedom equal to the number of parameters of interest. The parameters of interest are the vector $\s$, of length $N$, therefore 
\beq\label{chisqN}
t\overset{\textrm{AL}}{\sim}f_{\chi^2}(t;N)=\frac1{2^{N/2}\Gamma(N/2)} t^{N/2-1}\e^{-t/2}\,.
\eeq
The similarity of eq.~(\ref{PLR2}) with the maximum likelihood goodness-of-fit, which is also (in the AL limit) a $\chi^2$ with $N$ degrees of freedom, is apparent also in this respect.

The AL formula (\ref{chisqN}) is, as previously explained, essential for our construction. We must thus discuss in details the conditions for its validity. The AL is the one in which the data sample is large, a condition that we interprete in a restrictive manner by asking for $O_i\gg1$ in each bin.\footnote{A similar condition must hold for the auxiliary measurements that constrain $\n$.} Of course one cannot check the validity of the latter condition before the experiment is performed. Therefore for the design of the hypothesis test, prior to the experiment, we interprete the AL condition as $\M_i\gg1$ for the expected countings in each bin. The idea is that if $O_i$ will eventually turn out to be much different from $\M_i$, for instance if $\M_i\gg O_i \not\gg 1$, the hypothesis will be excluded anyhow and the possible unaccuracy of the AL formula in this regime will not be an issue. Of course $\M_i$ depends on $\n$, however the nuisance parameters are normally well constrained by the auxiliary measurements so that their variation within few sigmas around the central value $\n_0$ does not change $\M_i$ radically. The $\M_i\gg 1$ condition can thus be enforced at $\n=\n_0$. The $\M_i$'s also depend on the unknown additional signal components $\Delta_i$, and clearly we need the AL formula to hold for any $\Delta_i$. Since $\Delta_i\geq0$, asking for $\M_i\gg1$ at $\Delta_i=0$, i.e.
\beq\label{ALcond}
\S_i(\s,\n_0)+\B_i(\n_0)\gg1\,,
\eeq
is the necessary and sufficient condition for the validity of the AL formula in the entire $\D$-space. One might ask how large $\S_i+\B_i$ concretely needs to be for the AL to hold with good accuracy. In appendix~\ref{toy} we address this question quantitatively in a toy example and we find that values as small as $3$ are sufficient. Thousands of countings are expected in the mono-jet DM search we study in the next section, therefore the number of events is never an issue. The situation might be different for other applications of our method.

In order to construct a $\D$-independent hypothesis test we define $p_{\textrm{max}}$ like in eq.~(\ref{pvid}) and we use the AL formula (\ref{chisqN}) to express it, similarly to eq.~(\ref{pvAL}), as
\beq\label{tmindef}
\displaystyle
p_{\textrm{max}}\overset{{\rm{AL}}}{=}
\int_{t_{\rm{min}}}^\infty \hspace{-5pt}dt'\, f_{\chi^2}(t';N)\,,\;{\rm{where}}\;\;
t_{\textrm{min}}(\s;\o\,)=\underset{\Delta_i\geq0}{\textrm{\large{inf}}}\{t(\s,\D;\o\,)\}\,,
\eeq
with $t$ given in eq.~(\ref{PLR2}). Notice that $t$ is itself the result of a minimisation, with respect to $\n$. We must further minimise, with respect to $\D$, in order to obtain $t_{\textrm{min}}$. However it is convenient to invert the order of the two minimisations, because the one with respect to $\D$, if performed first, can be done analytically. The point is that the terms in the round brackets of eq.~(\ref{PLR2}) have a unique minimum (equal to zero) at $\M_i=O_i$ and monotonically increase (decrease) for $\M_i$ larger (smaller) than $O_i$. The $\Delta_i$'s, one for each bin, are positive shifts of $\M_i$ with respect to their minimal values $\S_i+\B_i$ (see eq.~(\ref{expnus})), therefore two possibilities are given. If $\S_i+\B_i\leq O_i$, i.e. if the observed counting over-fluctuates with respect to the prediction, a positive value of $\Delta_i$ exists that makes $\M_i=O_i$ and the absolute minimum is reached in the corresponding bin. If instead $\S_i+\B_i> O_i$, i.e. in an under-fluctuating bin, there is no way to reach $\M_i=O_i$ with a positive $\Delta_i$. The minimum is thus at $\Delta_i=0$, i.e. $\M_i=\S_i+B_i$. The position of the minimum in the $\D$-space and the corresponding formula for $t_{\textrm{min}}$ are 
\bea
&\displaystyle\left[\Delta_i(\s,\n)\right]_{\textrm{min}}\hspace{-4pt}=\left\{\begin{array}{lr}
\displaystyle O_i-\S_i(\n,\s\,)-\B_i(\n)\,,&{\textrm{if}}\; \S_i(\n,\s\,)+\B_i(\n)\leq O_i \\[6pt] 0\,,&{\textrm{if}}\; \S_i(\n,\s\,)+\B_i(\n)>O_i
\end{array}\right.\,,&\nonumber
\\ &\displaystyle{\mathlarger{{\mathlarger{\mathlarger{\Downarrow}}}}}&\nonumber\\
&\displaystyle t_{\textrm{min}}=
\underset{\n}{\textrm{\large{inf}}}\left[{\mathlarger{\sum\limits_{\{\S_i+\B_i>O_i\}}}}2\left(\S_i(\s,\n)+\B_i(\n)-O_i-O_i\log\frac{\S_i(\s,\n)+\B_i(\n)}{O_i}\right)-2\log\frac{{{L}}_\n}{{{L}}_{\n_0}}
\right]\,.&\label{tmin}
\eea

The result is trivial for its simplicity. Over-fluctuating bins do not contribute to $t_{\textrm{min}}$ because for them we set $\M_i=O_i$. The very simple reason for this is that $O_i\geq\S_i+\B_i$ is perfectly compatible with the hypothesis, the over-fluctuation of the observed being possibly due to some amount of additional signal. Under-fluctuating bins do instead contribute to $t_{\textrm{min}}$, and their contribution is conservatively evaluated at $\Delta_i=0$. Notice that in spite of the fact that they do not contribute to $t_{\textrm{min}}$, over-fluctuating bins are implicitly taken into account in our procedure. All the bins are indeed counted in the number of degrees of freedom $N$ of the $\chi^2$ distribution by which compatibility or incompatibility is established. An over-fluctuating (and thus perfectly compatible with the hypothesis, as explained) bin does not contribute to $t_{\textrm{min}}$ but contributes to $N$ and thus it increases the $p$-value making the model more compatible with data as it should.

Having defined the test for a generic model, with arbitrary parton-level cross-section in each bin, we can now straightforwardly restrict it to the model of interest, in which $\s$ depends on a single signal-strength parameter $\mu$ as $s_i(\mu)=\mu\,{\overline{s}}_i$.\footnote{We are assuming here that there is no interference between the signal and the SM, otherwise the dependence of $s$ on $\mu$ is not just a rescaling. We will see in section~\ref{inter} how to include interference in the discussion.} We just have to set, in eq.~(\ref{tmin})
\beq\label{truesig}
\S_i(\s,\n)=\S_i(\s(\mu),\n)=\mu \, \overline{{\S}}_i(\n)\,.
\eeq
We stress once again that this step does not require us to first compute $\vec\S$ as a function of $\s$ and next compute and plug-in $\s$ as a function of $\mu$. Everything can be done in one step by simulating the model with the benchmark value for its parameters, that corresponds to $\mu=1$, compute $\overline{\S}_i$ and eventually rescale by $\mu$. Once this is done, the substitution in eq.~(\ref{truesig}) makes $t_{\textrm{min}}$ become a function of $\mu$ (and of the observations) only and the exclusion limit on $\mu$, call it $\mu_{\textrm{exc}}$, is set by solving the equation
\beq\label{pexc}
p_{\textrm{exc.}}(\mu_{\textrm{exc}};\o\,)=
\int_{t_{\rm{min}}(\mu_{\textrm{exc}};\o)}^\infty \hspace{-5pt}dt'\, f_{\chi^2}(t';N)=1-\alpha\,,
\eeq
like for an ordinary hypothesis test.

We described the steps needed to apply our method in a rather pedantic manner, with the purpose of outlining that it does not require more work than the standard LHC limit-setting strategy. It requires signal simulations at different points of the $\n$-space, in order to assess the impact of nuisance on the signal, the determination of the backgrounds in control regions and the numerical minimisation over the nuisance parameters. All these steps are equally necessary for a standard analysis, which on the other hand cannot deal with additional signal. One additional technical difficulty might actually emerge in our case, due to the fact that whether or not one bin over-fluctuates, and should thus be included in the sum or not, might depend on the value of the nuisance parameters. This occurs when one of the bins has $O_i$ nearly equal to $\S_i+\B_i$ at the central value $\n=\n_0$ of the nuisance likelihood. In this case nuisance variations around the central value can turn the bin from under-fluctuating to over-fluctuating. The condition of summing over under-fluctuating bins in eq.~(\ref{tmin}) is practically implemented by a step function for each bin in front of the term in the round bracket. These step functions produce discontinuities where their arguments change sign. Actually we don't have discontinuities in the function or in its first derivative, because the term in the round bracket vanishes, together with its derivative, at the singular point $\S_i+\B_i=O_i$. Discontinuities appear in the second and higher derivatives. Such discontinuities might be impossible to treat for certain minimisation algorithms, in which case they should be regularised and this would result in a more demanding numerical procedure. We are not in the position to assess how severe the problem would actually be in a concrete LHC analysis (we easily dealt with it in the toy example in appendix~\ref{toy}), however we are confident that it can be circumvented. Nevertheless in the next section, in the process of simplifying our test, we will describe an approximate strategy that does not require minimising a non doubly-differentiable function.

\subsubsection{A first simplification}\label{fs}

A simple approximate version of our test is obtained by relying on three assumptions. The first one is that nuisance parameters ($\nu^{\hat{a}}$, $\hat{a}=1,\ldots,\kappa$) have Gaussian likelihood, i.e.
\beq\label{gausnuis}
{{L}}_{\vec\nu}=\frac1{(\sqrt{2\pi})^\kappa |V|}\e^{-\frac12(\n-\n_0)^tV^{-1}(\n-\n_0)}\,,
\eeq
with constant ($\n$-independent) covariance matrix $V$. The second assumption is that nuisance affects $\vec\S$ and $\vec\B$ mildly, so that it is legitimate to Taylor-expand them around $\n=\n_0$
\bea\label{Mexp}
\displaystyle
&&\M_i=\S_i+\B_i+\Delta_i\simeq \S_i^0+\B_i^0+\Delta_i+\delta\M_i\equiv \M_i^0+\delta\M_i\,,\\ \nonumber
\displaystyle
&&\delta\M_i=(\nu^{\hat{a}}\hspace{-2pt}-\hspace{-2pt}\nu_0^{\hat{a}})\partial_{\hat{a}}M_i^0=(\nu^{\hat{a}}\hspace{-2pt}-\hspace{-2pt}\nu_0^{\hat{a}})\left[
\frac{\partial\S_i}{\partial\nu^{\hat{a}}}\hspace{-2pt}+\hspace{-2pt}\frac{\partial\B_i}{\partial\nu^{\hat{a}}}
\right]_{\n=\n_0}\hspace{-10pt}\ll\M_i^0\,.
\eea
This is of course necessarily true if $\n$ is very close to $\n_0$. The assumption we make is that the linear expansion holds when $\n$ is few sigmas (in terms of the covariance matrix $V$) away from $\n_0$, which is the relevant region for the minimisation of eq.~(\ref{PLR2}). Eq.~(\ref{Mexp}) thus expresses the condition that the errors on the nuisance are small, in such a way that the few sigma region around $\n_0$ is small and the linear expansion is justified. Notice that few sigma variations of $\n$ resulting in a mild change of $\M$, i.e. $\delta\M\ll\M$, does not imply that the uncertainties on the nuisance are going to have a mild impact on the limit. Whether or not this is the case depends on how large or small $\M$ is in absolute value, which in turn determines the magnitude of the statistical error. In appendix~\ref{toy} we will come back to this point and to a quantitative assessment of how small the relative error due to nuisance needs to be for the validity of the approximation.

The third assumption is that the countings are approximately Gaussian distributed, with mean and variance $\M_i$, so that the $\D$-dependent test statistic in eq.~(\ref{PLR2}) becomes
\beq\label{PLR2gauss}
\displaystyle
t(\s,\D;\o\,)=\underset{\n}{\textrm{\large{inf}}}\left\{\sum\limits_{i=1}^N\frac{(\M_i-O_i)^2}{\M_i}+(\n-\n_0)^t V^{-1}(\n-\n_0)
\right\}
\,.
\eeq
Poisson countings automatically become Gaussian in the large-$\M_i$ limit we already assumed in eq.~(\ref{ALcond}) in order to obtain a $\chi^2$-distributed test statistic as in eq.~(\ref{chisqN}). However the Gaussian approximation for the Poisson distribution becomes accurate enough only for values of $\M_i$ which are larger (above around few tenths, see appendix~\ref{toy}) than those needed for the validity of eq.~(\ref{chisqN}). 

Few straightforward manipulations, reported for completeness in appendix~\ref{appB}, allow us to compute eq.~(\ref{PLR2gauss}) analytically in the $\delta\M_i\ll\M_i$ limit of eq.~(\ref{Mexp}). The result is the familiar $\chi^2$ formula
\beq\label{chisqt}
\displaystyle
t\simeq\chi^2=\sum\limits_{i,j=1}^{N}(\M_i^0-O_i)(\Sigma_{\textrm{tot}}^{-1})_{ij}(\M_j^0-O_j)\,,\;\;\;\;\;\Sigma_{\textrm{tot}}=\Sigma_\M+\Sigma_\nu\,,
\eeq
where the total covariance matrix $\Sigma_{\textrm{tot}}$ is the sum of statistical ($\Sigma_\M$) and systematical ($\Sigma_\nu$) errors. Let us discuss them in turn. The statistical error is the variance of the Poisson distribution in each bin, evaluated at the central value of the nuisance parameters $\n=\n_0$, i.e.
\beq\displaystyle\label{covse}
(\Sigma_\M)_{ij}=\M_i^0\delta_{ij}\,.
\eeq
Obviously it is diagonal, since statistical errors are uncorrelated. The systematical component depends on the nuisance covariance matrix $V$ and on how sensitive the $\M_i$'s are to nuisance departures from the central value. It is given by a formula
\beq\displaystyle\label{covsis}
(\Sigma_\nu)_{ij}=\sum\limits_{\hat{a},\hat{b}=1}^{\kappa}\partial_{\hat{a}}\M_i^0V_{\hat{a}\hat{b}}\partial_{\hat{b}}\M_j^0\,,
\eeq
which seems, at first sight, complicated to apply in a concrete analysis where often neither $V$ nor the functional dependence of $\vec\M$ on $\n$ are known explicitly. However by the standard error propagation formula we can rewrite it as
\beq\displaystyle\label{covsis1}
(\Sigma_{\nu})_{ij}=\int\hspace{-2pt} d^{\kappa}\nu\, {\mathcal{L}}_\n\cdot (\M_i(\n)-\M_i^0)(\M_j(\n)-\M_j^0)=E\left[(\M_i-\M_i^0)(\M_j-\M_j^0)\right]
\,,
\eeq
where ``$E$'' denotes the expectation value over the nuisance parameters, treated here as random variables with p.d.f. ${\mathcal{L}}_\n$. By sampling the $\n$-space with weight ${\mathcal{L}}_\n$, computing $\vec\M$  by means of simulations and taking averages, $\Sigma_\nu$ is easily obtained with no need of determining $V$ and $\partial_{\hat{a}}\M_i$ as an intermediate step.\footnote{A similar logic can be applied to $\M^0_i$, which is equal to $E[\M_i]$. However the approach of first computing $\n_0$ as the maximum of ${\mathcal{L}}_\n$ and next using it to obtain the $\M^0_i$'s is probably more convenient.}

Having expressed $t(\s,\D;\o\,)$ analytically by the $\chi^2$ formula, we should now minimise it over the additional signal components $\Delta_i\geq0$, as prescribed by eq.~(\ref{tmindef}). While this minimisation will in general be performed numerically, it is important to get an analytical understanding of the dependence of $t$ on $\D$, the way it emerges from the various elements that compose the $\chi^2$ formula~(\ref{chisqt}). The first element is $\M_i^0$, which is given by
\beq\displaystyle\label{mi0}
\M_i^0=\mu\,\overline{{\S}}_i(\n_0)+\B_i(\n_0)+\Delta_i\,,
\eeq
having made use of eq.~(\ref{truesig}) to restrict the signal to the one predicted by our model in terms of the unique signal-strength parameter $\mu$. The $\M_i$'s depend on $\D$, but in a trivial (additive) manner. The terms $\overline{{\S}}_i^{\raisebox{-3pt}{\scriptsize{0}}}$ and $\B_i^0$ are of course independent of $\D$ and thus they can be computed once and for all by setting the nuisance parameters to their central values. For the nuisance that correspond to instrumental effects (e.g., trigger or reconstruction), this amounts to perform one single signal and background simulation with the nuisance set to their nominal values. For nuisance associated with backgrounds, it amounts to compute the background at the central value obtained from the fit in the control regions. The dependence on $\mu$ of the signal is included by rescaling the benchmark simulation. The second element we need in the $\chi^2$ formula is $\Sigma_\nu$ which, importantly enough, is independent of $\D$. This being the case is obvious from its definition, but also from rewriting eq.~(\ref{covsis1}) explicitly as
\bea\displaystyle\label{covsis2}
(\Sigma_{\nu})_{ij}=&&\hspace{-15pt}\mu^2\,E[(\overline{{\S}}_i-\overline{{\S}}_i^{\raisebox{-3pt}{\scriptsize{0}}})(\overline{{\S}}_j-\overline{{\S}}_j^{\raisebox{-3pt}{\scriptsize{0}}})]
+\mu\,
E[(\overline{{\S}}_i-\overline{{\S}}_i^{\raisebox{-3pt}{\scriptsize{0}}})(B_j-B_j^0)+i\leftrightarrow j]
\nonumber\\
\displaystyle
&&\hspace{-15pt}+ E\left[(\B_i-\B_i^0)(\B_j-\B_j^0)\right]
\,.
\eea
This equation allows us to compute $\Sigma_{\nu}$ as a function of the signal strength parameter by sampling the nuisance space as previously explained. Together with eq.~(\ref{mi0}), it provides us with $t$ as function of $\mu$ and $\D$. After minimising over $\D$ we obtain ${t_{\rm{min}}(\mu;\o)}$ and in turn $\mu_{\textrm{exc}}$ by solving eq.~(\ref{pexc}).

The minimisation of $t$ will have to be performed numerically, but this does not pose any conceptual difficulty as the $\chi^2$ is an infinitely differential function of $\D$. The procedure can however be rather slow because each function call requires the inversion of $\Sigma_{\textrm{tot}}$, which depends on $\D$ through $\Sigma_\M$. A considerable simplification is obtained by replacing eq.~(\ref{chisqt}) with the ``modified'' $\chi^2$ formula
\beq\label{chisqtmod}
\displaystyle
t\simeq\chi^2_{\textrm{mod}}=\sum\limits_{i,j=1}^{N}(\M_i^0-O_i)(\Sigma_{\textrm{tot}}^{-1})_{ij}(\M_j^0-O_j)\,,\;\;\;\;\;\Sigma_{\textrm{tot}}=\Sigma_O+\Sigma_\nu\,,
\eeq
where $\Sigma_O$ is the statistical error matrix computed with the observed countings rather than with the expected ones. The logic behind this simplification is that $\M_i$ cannot be much different from $O_i$, in the limit of high statistic, for configurations that are not trivially excluded. The modified $\chi^2$ is a simple quadratic function of $\D$, which can minimised either analytically (though with some complication due to the condition $\Delta_i\geq0$) or numerically with a fast procedure since no $\D$-dependent matrix inversion is involved in the evaluation of the function. The modified $\chi^2$ is known to be a poor approximation of the exact one, that requires large statistics to become accurate. We verify this fact explicitly with our toy example in appendix~\ref{toy}. However eq.~(\ref{chisqtmod}) is useful also when the statistics is not large in order to get a first estimate of the location of the minimum in the $\D$-space, to be used as a convenient starting point for the minimisation of the exact $\chi^2$ formula.

\subsubsection{Simplified likelihood}\label{simL}

The limit-setting strategies outlined above require complete knowledge of the data and full control of the sources of systematic uncertainty. Thus they must be carried on internally by the experimental collaborations. Alternative approaches, in which the limit is set externally to the collaborations, are highly desirable because they avoid the presentation of the experimental results being committed to a specific (though quite general, like an EFT) new physics model. The way to proceed is, ideally, rather obvious. The experiments should select a set of quantities calculable with ease by theorists that effectively parametrise a generic signal hypothesis, and report the likelihood as a function of these quantities. If the likelihood is Gaussian, it is of course sufficient to report central values and covariance. In the case of DM, valid quantities to be considered are certainly the parton-level DM plus $X$ production cross-sections in each $\MET$ bin, call them $\sigma_i$.\footnote{These should not be confused with the $s_i$'s we defined in section~\ref{full}. The latter are the EFT contribution to the cross-section, while $\sigma_i$ is the total cross-section emerging from the sum of the EFT and of the unknown UV contribution.} If $L(\vec\sigma)$ was given, one could straightforwardly apply the logic we outlined in section~\ref{thp} and carry on limit-setting with no need of further experimental information. One would do so by adopting for eq.~(\ref{lik0}) a slightly different interpretation than the one of sections~\ref{full} and \ref{fs}. The signal $\vec\S$ would be the contribution to $\vec\sigma$ (call it $\vec\sigma_{\textrm{\tiny{EFT}}}$) coming from the EFT truncated by the $E_{\rm{cm}}<M_{\textrm{cut}}$ restriction. $\D$ would be the additional parton-level DM production cross-section rather then the additional detector-level signal as in sections~\ref{full} and \ref{fs}. The test statistic would be defined as the Log Likelihood Ratio and would be distributed as a $\chi^2$ with $N$ degrees of freedom in the AL. After computing $\vec\sigma_{\textrm{\tiny{EFT}}}=\mu\,\vec{\overline\sigma}_{\textrm{\tiny{EFT}}}$ at parton level, plugging in $t$ and maximising over $\D$ we would end up with eq.~(\ref{pexc}) and use it to set the limit.

The above procedure has limitations, the first one being that it requires quite a bit of additional experimental work. Computing $L(\vec\sigma)$ requires taking the benchmark model simulation, rescaling it with one signal-strength parameter for each histogram bin and studying the dependence of the likelihood on these parameters. The second difficulty is that it exposes us to errors that is difficult to quantify, associated with the fact that other kinematical distributions (other than $\MET$) of the benchmark simulation might be different from those of the EFT signal to which the analysis will eventually be applied. Such difference might affect the experimental efficiencies and produce a non-accurate result. Even more worrisome is the fact that the kinematical distributions of the additional signal are completely unknown. Notice that this is not an issue with the approaches of sections~\ref{full} and \ref{fs} because $\D$ there represents the additional signal yield, whose connection with the additional cross-section needs not to be specified.

Another approach to limit reinterpretation, that does not suffer of the issues above, is based on a ``Simplified Likelihood'' \cite{Collaboration:2242860,Fichet:2016gvx}. The way it works is nicely described by looking at eq.~(\ref{covsis2}). Assume the approximations we made in section~\ref{fs} are justified, so that $t$ becomes the $\chi^2$ in eq.~(\ref{chisqt}), with $\Sigma_\nu$ as in eq.~(\ref{covsis2}). Suppose also, and this needs to be checked case-by-case, that none of the relevant sources of nuisance affect the signal and the background at the same time, or that nuisance effects on the signal are negligible. In the mono-jet DM search we discuss in section~\ref{DMEFTLIM} this is quite a reasonable approximation since data-driven background estimate is the major source of uncertainty. In the hypothesis that signal and background nuisance parameters are disjoint, the mixed terms in eq.~(\ref{covsis2}) drop and one obtains
\beq\displaystyle\label{covsis3}
(\Sigma_{\nu})_{ij}=\mu^2\,E[(\overline{{\S}}_i-\overline{{\S}}_i^{\raisebox{-3pt}{\scriptsize{0}}})(\overline{{\S}}_j-\overline{{\S}}_j^{\raisebox{-3pt}{\scriptsize{0}}})]
+ E\left[(\B_i-\B_i^0)(\B_j-\B_j^0)\right]\equiv\mu^2(\Sigma_{S})_{ij}+(\Sigma_{B})_{ij}\,.
\eeq
The simplified likelihood proposal \cite{Collaboration:2242860,Fichet:2016gvx} is that the experimental collaborations report $\Sigma_{B}$, leaving to theorist the determination of $\Sigma_{S}$, if needed. The central value of the background in each bin, $\B_i^0$, will also be reported, while the evaluation of $\overline{{\S}}_i^{\raisebox{-3pt}{\scriptsize{0}}}$ is again left to theory estimate. The potential limitation of this method is that it relies on an accurate simulation of the detector-level signal $\overline{{\S}}_i^{\raisebox{-3pt}{\scriptsize{0}}}$ and, even worse, of $\Sigma_{S}$, if not negligible. In section~\ref{DMEFTLIM} we will apply the simplified likelihood method to the CMS mono-jet search, and we will argue that it should be reasonably accurate for our purpose. Validation from a full-fledged EFT experimental analysis would however be welcome.

\subsubsection{Cross-section measurements}
The last case which is worth discussing is the one in which the experimental collaborations report, as the result of the analysis, measurements of unfolded parton-level differential cross-sections. When accurate theoretical predictions of the SM backgrounds are available, this makes setting limits on the EFT a very simple task. Clearly this approach is not applicable to DM searches, where the background cannot be predicted.\footnote{In principle, one might consider providing both a measurement of the $\MET$ distribution and a measurement of the SM background from control regions. Addressing the feasibility of this strategy goes beyond the purpose of the present article.} However it could be useful for other EFT studies such as the ones proposed in refs.~\cite{Farina:2016rws,deBlas:2013qqa}. The logic outlined in section~\ref{thp} straightforwardly applies to cross-section measurements. The signal $\vec\S$ in eq.~(\ref{lik0}) represents now the parton-level EFT cross-sections ${\vec\sigma}^{\textrm{\tiny{EFT}}}$, in $N$ bins, computed within the EFT with the habitual $E_{{\rm{cm}}}<{M_{\rm{cut}}}$ restriction.  $\D$ is the additional signal cross-section from reactions above the EFT cutoff and the observations $\o$ are the measured cross-sections ${\vec\sigma}^{\textrm{m}}$. The test statistic is just the $\chi^2$
\beq
\displaystyle
t(\vec\S,\D;\o\,)=\chi^2=\sum\limits_{i,j=1}^N({\vec\sigma}^{\textrm{\tiny{EFT}}}\hspace{-2pt}+\D+\vec{b}-{\vec\sigma}^{\textrm{m}})_i\Sigma^{-1}_{ij}
({\vec\sigma}^{\textrm{\tiny{EFT}}}\hspace{-2pt}+\D+\vec{b}-{\vec\sigma}^{\textrm{m}})_j
\,,
\eeq
where $\Sigma$ is the covariance matrix of the measurements. The SM background $\vec{b}$ is taken here to be predicted with infinite accuracy, but of course it would not be hard to include the theory uncertainties on the background prediction in the $\chi^2$ formula by proceeding like in the previous section. The test statistic is distributed as a $\chi^2$ with $N$ d.o.f., and it is very easy to maximise over $\Delta_i\geq0$, being just a quadratic function. The limit is thus set as a trivial application of eq.~(\ref{pvAL}).

\subsubsection{Dealing with interference}\label{inter}

Until now we described our limit-setting strategy having implicitly (or even explicitly, see eq.~(\ref{truesig})) in mind the case in which the EFT produces a phenomenon (e.g., DM production) that does not occur in the SM. If this is the case, no quantum mechanical interference is present between the SM and the EFT Feynman diagrams and  a clear distinction can be made between the ``background'' and the ``signal'', to be further split into the proper EFT signal and the unknown additional signal from high-energy reactions. Specifically, in DM mono-$X$ searches the background is the complete SM contribution to the $\MET$ distribution, emerging from SM processes with arbitrarily high $E_{\rm{cm}}$. The signal is instead DM production from $E_{\rm{cm}}<M_{\textrm{cut}}$, to be computed starting from the EFT diagrams evaluated with the $E_{\rm{cm}}<M_{\textrm{cut}}$ restriction on the phase space. The additional signal is the UV contribution to DM production.

Other interesting EFT's are those that describe BSM effects in the EW plus Higgs sector. These EFT's normally produce BSM corrections to processes that do occur also in the SM, and as such they do interfere with the SM diagrams. In the presence of interference our method should be applied as follows. The background must be interpreted as the sum of all SM processes that do not interfere with the EFT because they are characterised by a different parton-level final or initial state. We can call it the ``ideally reducible'' background, as one might get rid of it by an ideal detector capable to reconstruct all the particles (including, say, the neutrinos) with infinite accuracy. The ``signal'' is all the rest. It contains the pure EFT contribution, the SM-EFT interference and the SM prediction for the relevant final state. This is further split into ``proper'' low-energy EFT signal with the $E_{\rm{cm}}<M_{\textrm{cut}}$ restriction, plus ``additional signal'' from the UV. Specifically this means that the signal also contains SM terms (namely, the square of SM diagrams), which are also truncated by $E_{\rm{cm}}<M_{\textrm{cut}}$. Similarly, the unknown additional signal is the physical contribution to the distribution that comes from $E_{\rm{cm}}>M_{\textrm{cut}}$ processes and as such it also includes high energy SM contributions. Notice that with this definition the additional signal $\Delta_i$ is positive (and the signal as well) because the distinction between the EFT signal and the additional one is made on a physical (kinematical) basis. This would not be the case if we had left the SM out of the truncation and regarded it as part of the background like was done in ref.~\cite{Falkowski:2016cxu}.

With this interpretation in mind one can easily go through the previous sections and check that all the consideration we made remain valid. Our method can thus  straightforwardly deal with interference, eq.~(\ref{truesig}) being the only formula that concretely needs to be modified. It assumes the signal being proportional to the signal-strength parameter $\mu$, which is interpreted as the total signal cross-section. In the presence of interference $\mu$ should be better viewed as the coefficient of the EFT operator and eq.~(\ref{truesig}) takes the form
\beq\label{truesigint}
\S_i(\s,\n)=\S_i(\s(\mu),\n)=\mu^2 \overline{{\S}}_i^{(2)}(\n)+\mu\, \overline{{\S}}_i^{(1)}(\n)+ \overline{{\S}}_i^{(0)}(\n)\,.
\eeq
The three terms correspond respectively to the square of the EFT diagram, to the interference and to the SM contribution. All the previous formulas that rely on eq.~(\ref{truesig}) can be easily modified according to eq.~(\ref{truesigint}).

\subsubsection{Why not a signal-strength-based test ?}\label{whynot}

We extensively discussed in section~\ref{full} that our test is slightly ``unusual'' because it is first constructed for a generic model in which the signal cross-section in each bin is a free parameter, and later restricted to the model of interest where the signal-strength $\mu$ is the only parameter. Ordinary LHC tests are instead ``signal-strength-based'' from the very beginning, namely they are constructed having in mind that $\mu$ is the only parameter of the probability model. Focussing for simplicity on the case in which the interference is absent, the restricted signal is the one in eq.~(\ref{truesig}). In the construction of signal-strength-based tests, this restriction is applied already in eq.~(\ref{PLR1}), both in the numerator and the denominator of the likelihood ratio. This makes no difference for the numerator, where we also make use of eq.~(\ref{truesig}), eventually, but it is a big change for the denominator, where now the maximisation is performed over $\mu$ only rather than on $\s$. The signal-strength-based test has typically a stronger expected limit than the one we used, one might thus wander why we did not take this direction in our construction. The point is that what the signal-strength-based test actually does is comparing the hypothesis we want to test with the most favourable hypothesis (i.e., the maximal $\mu$) that is present in the restricted set defined by eq.~(\ref{truesig}). If all the hypotheses in the set are far from the data, it returns an artificially high $p$-value. This is precisely the situation we encounter if we consider very large (positive) values for the additional signal $\Delta_i$, much above the $\mu \, \overline{{\S}}_i$ contribution from the EFT and the observed $O_i$. In this configuration, the total expected is much larger than the observed and the model is in tension with observations for any value of $\mu$. Moreover the likelihood is nearly constant in $\mu$ because $\mu \, \overline{{\S}}_i\ll\Delta_i$, therefore the likelihood divided by its maximum is nearly $1$, i.e. $t\simeq0$, and the signal-strength-based test returns perfect compatibility. Thus a signal-strength-based test cannot be applied in the presence of additional signal because the minimisation of $t$ over $\D$ would always return $t_{\textrm{min}}=0$, with the minimum reached at $\D\rightarrow\infty$.

\section{CMS mono-jet search reinterpretation}\label{DMEFTLIM}

As a simple example of DM EFT, we consider DM being a Majorana particle $\chi$ in the singlet of the EW group. If DM (with mass $\mdm$) is the lightest particle of its sector, and for energies below the mass ``$M_{\textrm{med}}$'' of the other new particles (where $M_{\textrm{med}}\gg\mdm$), its interactions with the quarks are universally described by a set of $d=6$ effective operators 
\begin{equation}\displaystyle\label{opmaj}
 \mathcal{L}_{\textrm{int}} = \frac{1}{M_*^2} \sum_i c_i O_i\,,
\end{equation}
classified in ref.~\cite{Goodman:2010ku}. The effective operator Wilson coefficients are conveniently parametrised in terms of dimensionless parameters $c_i$ and of an overall interaction scale $M_*$. The latter scale should not be confused with the EFT cutoff $M_{\textrm{cut}}$ or with the mass $M_{\textrm{med}}$ of the heavy particles mediating the quark-DM interaction. In order to avoid confusion it would have been convenient to trade it for a parameter $G_*\equiv1/M_*^2$, analog to the Fermi constant. We will nevertheless adopt the standard notation and use $M_*$ instead. A comprehensive study of all the operators in eq.~(\ref{opmaj}) would be interesting, and straightforward with our methodology. However for the sake of simplicity we restrict here to a single axial-axial operator
\begin{equation}\displaystyle
\label{effectiveoperator}
\mathcal{L}_{\textrm{int}} = - \frac{1}{M_*^2} (\overline{X} \gamma^{\mu} \gamma^5 X)(\sum_q \bar{q} \gamma_{\mu} \gamma^5 q)\,,
\end{equation}
where $q$ runs over the six species of SM quarks. This specific operator is not particularly motivated from a BSM perspective, still it has been extensively studied in the literature and several mediator models have been proposed for its microscopic origin. References can be found in \cite{Racco:2015dxa}.

The EFT we will study is thus endowed with a $3$-dimensional parameter space, the parameters being the DM mass $\mdm$, the interaction scale $M_*$ and the cutoff of the EFT $M_{\textrm{cut}}$. The latter should be regarded and treated, for all practical purposes, as one of the free parameters of the EFT \cite{Racco:2015dxa}. Notice that there is no way to define rigorously $M_{\textrm{cut}}$, nor to get a hint of its value, before the UV-completion is specified. This is why it is important to treat it as a free parameter and show how the limits change as a function of $M_{\textrm{cut}}$. Qualitatively, $M_{\textrm{cut}}$ is of the order of the mass of the mediator particles, $M_{\textrm{med}}$, but it does not necessarily coincides with that. $M_{\textrm{cut}}$ is the maximal energy at which the EFT predictions resemble those of the UV model, and thus it will typically have to be taken slightly below $M_{\textrm{med}}$. Deciding how much below, and thus choosing the relevant $M_{\textrm{cut}}$ exclusion contour among the ones we will draw, is left to model-builders aimed at using our results to set limits on their specific UV model. On top of displaying limits in the $\mdm$-$M_*$ plane at fixed $M_{\textrm{cut}}$, one can also visualise exclusions at fixed $g_*$, with $g_*$ defined as 
\begin{equation}\label{gstar}\displaystyle
G_* = \frac{1}{M_*^2} = \frac{g_*^2}{M_{\textrm{cut}}^2}\,.
\end{equation}
The advantage is that $g_*$ ranges in a finite domain because basic perturbativity considerations require $g_*\lesssim4\pi$ and $g_*\not\ll1$ is expected for WIMP-like DM \cite{Racco:2015dxa}.

In what follows, limits are derived on this specific EFT by re-interpreting the CMS $13$~TeV mono-jet search \cite{Sirunyan:2017hci}, for which the Simplified Likelihood is available, as an application of the methodology developed in the previous section. 

\subsection{Signal simulation}

\begin{table}
\begin{center}
\begin{tabular}{l c c c c c c c c}
$\bm{m_{\rm\bf{\tiny{DM}}}}$ {\bf{[GeV]}} & 1 & 100 & 200 & 400 & 600 & 800 &1000  \\
\midrule
$\bm{\bar{\sigma}}$ \textbf{[pb]} & $25$ & $16$ & $11$ & $5.1$ & $2.6$ & $1.3$ & $0.68$ \\
\bottomrule
\end{tabular}
\end{center}
\caption{Total DM pair-production cross-section, for $M_*=1$~TeV at the $13$~TeV LHC.\label{table1}}
\end{table}

We simulated the DM production signal with {\sc{MadGraph 5}} \cite{Alwall:2011uj}, 
interfaced with {\sc{PYTHIA 6}} \cite{Sjostrand:2006za} for showering and hadronization and with {\sc{Delphes}} \cite{deFavereau:2013fsa} for the simulation of the detector response. DM pair-production $pp\rightarrow\chi\chi$ is simulated with up to two parton-level jets in the final state and the resulting event samples are combined by the MLM showering/parton-level matching implemented in {\sc{MadGraph}}. The effective operator scale has been set to $M_*=1$~TeV and seven simulations have been performed at the $\mdm$ points listed in table~\ref{table1}. The values reported in the table represent the total production cross-section, inclusive in the number of jets, with no cuts.

In a conventional BSM search, the only cuts to be imposed on the generated event sample would be those related with trigger, acceptance and selection that define the search region employed in the experimental analysis.\footnote{\label{SRC}
In our case \cite{Sirunyan:2017hci}, the cut that define the signal region are $\MET>200$~GeV, $p_T>100$~GeV and $|\eta|<2.5$ for the leading jet in $p_T$.} For an EFT-based signal instead, the $E_{\rm{cm}}<M_{\textrm{cut}}$ restriction must also be put in place and the question arises of how the center of mass energy $E_{\rm{cm}}$ should be concretely defined. If the signal was uniquely associated to a single parton-level hard process we would naturally define $E_{\rm{cm}}$ as the center of mass energy of that partonic process. If for instance the signal was entirely produced by $pp\rightarrow\chi\chi j$, $E_{\rm{cm}}$ would be the total DM-pair plus parton-level jet invariant mass. Extra low-$p_T$ jets emitted by parton showering must be excluded from the calculation of $E_{\rm{cm}}$ because those emissions are low virtuality QCD processes, whose occurrence does not invalidate the accuracy of the EFT prediction. While $pp\rightarrow\chi\chi j$ is indeed the dominant process in our sample, after restricting it to the mono-jet search region, the contribution from other partonic configurations is not completely negligible and a more refined definition of $E_{\rm{cm}}$ is needed. This is constructed by noticing that the purpose of matching algorithms is precisely to distinguish hard jet emissions, that do contribute to $E_{\rm{cm}}$, from soft ones that do not. The definition of $E_{\rm{cm}}$ is particularly straightforward to construct within the MLM matching algorithm \cite{Alwall:2007fs}, because this algorithm removes all the events in which soft QCD emissions are generated at parton level, so that the soft emissions are exclusively generated by showering in all the events that compose the final sample. The parton-level configuration that produced each event, stored in the output file, is thus the proper hard process to be associated with the event, and $E_{\rm{cm}}$ is computed out of that. Once the detailed implementation of the algorithm (MLM Kt-jet as implemented in {\sc{MadGraph 5}}, in our case) and its parameters (``xqcut = 30 GeV'', ``ptj=xqcut'', ``etaj = 7'', ``maxjetflavor = 5'' and ``QCUT = 100'') are specified, this definition of $E_{\rm{cm}}$ is unambiguous and fully reproducible.

The $E_{\rm{cm}}$ definition given above, based on the MLM method, is theoretically accurate and easy to implement, it is thus the one we will use in what follows. However it is interesting to compare it with alternative reasonable definitions, given below
\begin{itemize}
\item \textbf{Leading jet} : The total invariant mass of the system composed of the two DM particles and of the ${p_{T}}$-leading jet after the jet reconstruction.
\item \textbf{Multiple jets} : The jets of an event are selected and ordered in $p_{T}$. Then the leading jet's $4$-momentum is summed to the one of the DM particles. If the transverse component of the total $4$-momentum calculated this way is at least $90\%$ of the event's $\MET$, $E_{\rm{cm}}$  is calculated as the total $4$-momentum invariant mass. If this does not happen, the second $p_{T}$-ordered jet is considered, its $4$-momentum is summed to the previous one and then the transverse component is evaluated again. The procedure goes on until the $\MET$ value is balanced at $90\%$ percent by the jets.
\end{itemize}
\begin{figure}
\centering
\includegraphics[width= 8.5cm]{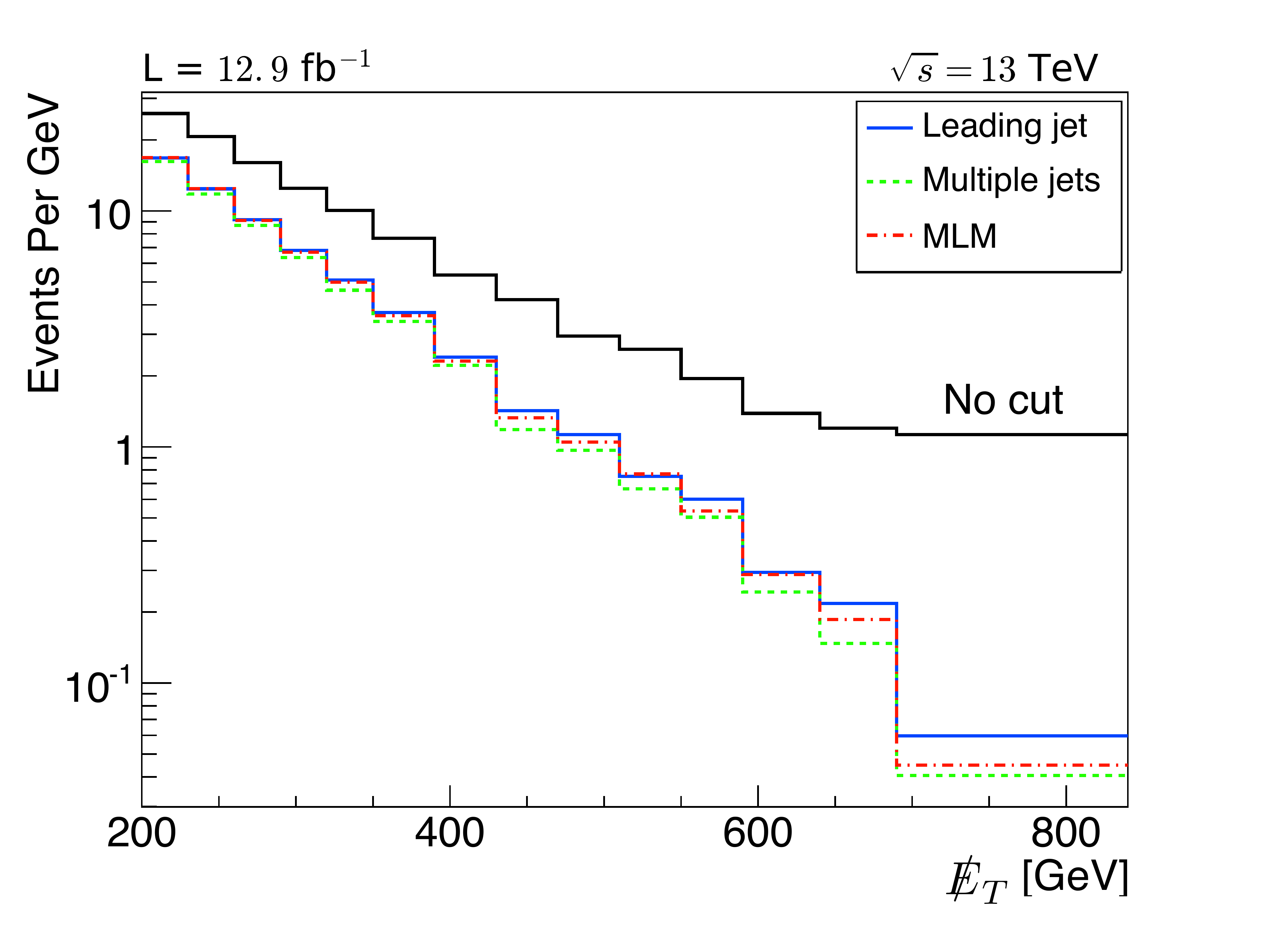}\includegraphics[width= 8.5cm]{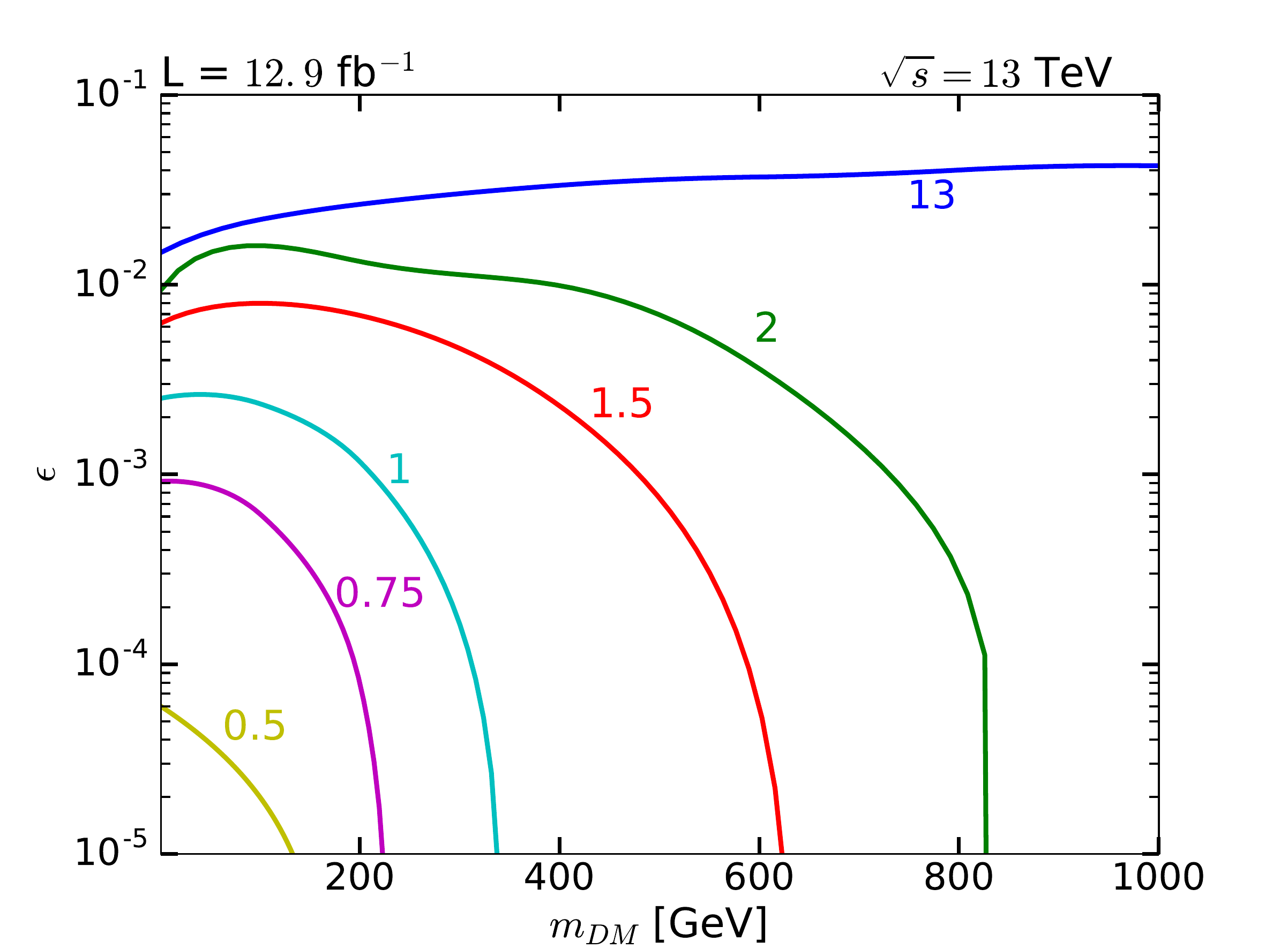}
\caption{Left Panel: comparison among the three definitions of $E_{\rm{cm}}$ described in the text. The distribution without $E_{\rm{cm}}$ cut is also shown. Right Panel" The efficiency $\epsilon$ as a function of $\mdm$ at different $M_{\textrm{cut}}$ values.} \label{metodi}
\end{figure}
On the left panel of figure \ref{metodi}, the $\MET$ distribution is shown for $\mdm = 200$~GeV, after the cuts specified in footnote~\ref{SRC} and the $E_{\rm{cm}}<M_{\textrm{cut}}=2$~TeV restriction, with the three different definitions of $E_{\rm{cm}}$ given above. The distribution without any $E_{\rm{cm}}$ cut is also shown for comparison. The three definitions give quite similar results, showing however appreciable differences. The ``Leading jet'' definition tends to overestimate the EFT signal because it does not count hard extra jets emission in the calculation of $E_{\rm{cm}}$, thus producing a lower $E_{\rm{cm}}$ and in turn a higher signal. The ``Multiple jets'' instead underestimates the EFT signal, showing that showering emissions give a significant contribution to the transverse momentum unbalance of the event. 

After the $E_{\rm{cm}}<M_{\textrm{cut}}$ restriction and with the cuts in footnote~\ref{SRC}, the total EFT signal cross-section takes the form
\begin{equation}\displaystyle
\label{crossxEFT}
\sigma_{\rm{\tiny{EFT}}} (M_*, \mdm, M_{\textrm{cut}}) = \left [ \frac{1\,{\textrm{TeV}}}{M_*} \right ]^4 \cdot \overline{\sigma} (\mdm) \cdot \epsilon (\mdm,  M_{\textrm{cut}})\,, 
\end{equation}
with $\overline{\sigma}$ as in table~\ref{table1}. The scaling with the fourth power of $1/M_*$ is dictated by the one of the effective operator coefficient. The efficiency $\epsilon$ is evaluated on the seven $\mdm$ points in table~\ref{table1} and for seven values of $M_{\textrm{cut}}$:
\begin{center}
\begin{tabular}{c c c c c c c}
\multicolumn{7}{c}{$\mathbf{M_{\textrm{\bf{cut}}}}$ {\bf{[GeV]}}}\\ 
\midrule
$500$ & $750$ & $1000$ & $1500$ & $2000$ & $4000$ & $13000$ \\ 
\bottomrule
\end{tabular} 
\end{center}
The highest value is of course equivalent to $M_{\textrm{{{cut}}}}=\infty$, namely it assumes that the EFT is valid at all the energy scale the LHC can probe. 

It is interesting to inspect the dependence of $\epsilon$ on $\mdm$ and $M_{\textrm{{{cut}}}}$, shown on the right panel of figure~\ref{metodi}. In the Naive EFT, the efficiency monotonically increases with $\mdm$ because heavier DM is produced in more energetic reactions, for which it is easier to produce large enough $\MET$ and jet $p_T$ to contribute to the signal region. The efficiency instead decreases with $\mdm$ when the $E_{\rm{cm}}$ cut is in place, because asking for $E_{\rm{cm}}<M_{\textrm{cut}}$ forbids hard reactions and reduces the allowed phase space. The efficiency goes exactly to zero above the kinematical threshold
\begin{equation}\displaystyle\label{thr}
(\mdm^{\textrm{max}})^2 = \frac{M_{\rm{cut}}^2}{4} \left({1 - 2 \frac{\MET}{M_{\rm{cut}}}}\right)\,.
\end{equation}

\subsection{Setting limits}

The CMS mono-jet analysis \cite{Sirunyan:2017hci} contains all the information we need to set limits on the EFT as a straightforward application of section~\ref{simL}. Namely it reports (see table~1 and figure~14 of \cite{Sirunyan:2017hci}) the observed countings $O_i$, the expected backgrounds $\B_i^0$ with their covariance matrix $\Sigma_{B}$ (see eq.~(\ref{covsis3})), in $22$ $\MET$ bins. However we considered the number of countings in the high-$\MET$ bins slightly too small for a safe application of the $\chi^2$ formula (see appendix~\ref{toy}), therefore we aggregated bins number $14$ to $16$ and $17$ to $22$ in two single bins and we employed an $N=15$ bins histogram to set the limit. The covariance matrix for the aggregated bins is obtained as in ref.~\cite{Collaboration:2242860}. The simplified likelihood approach prevents a full-fledged treatment of nuisance, but still it allows us to introduce, through $\Sigma_S$, nuisance parameters that affect only the signal such as the luminosity and the trigger uncertainties. We verified that those effects do not change our limit appreciably, therefore we set $\Sigma_S=0$ in what follows.

\begin{figure} 
	\centering
	\begin{minipage}[b]{8cm}
	\centering
	\includegraphics[width=8cm]{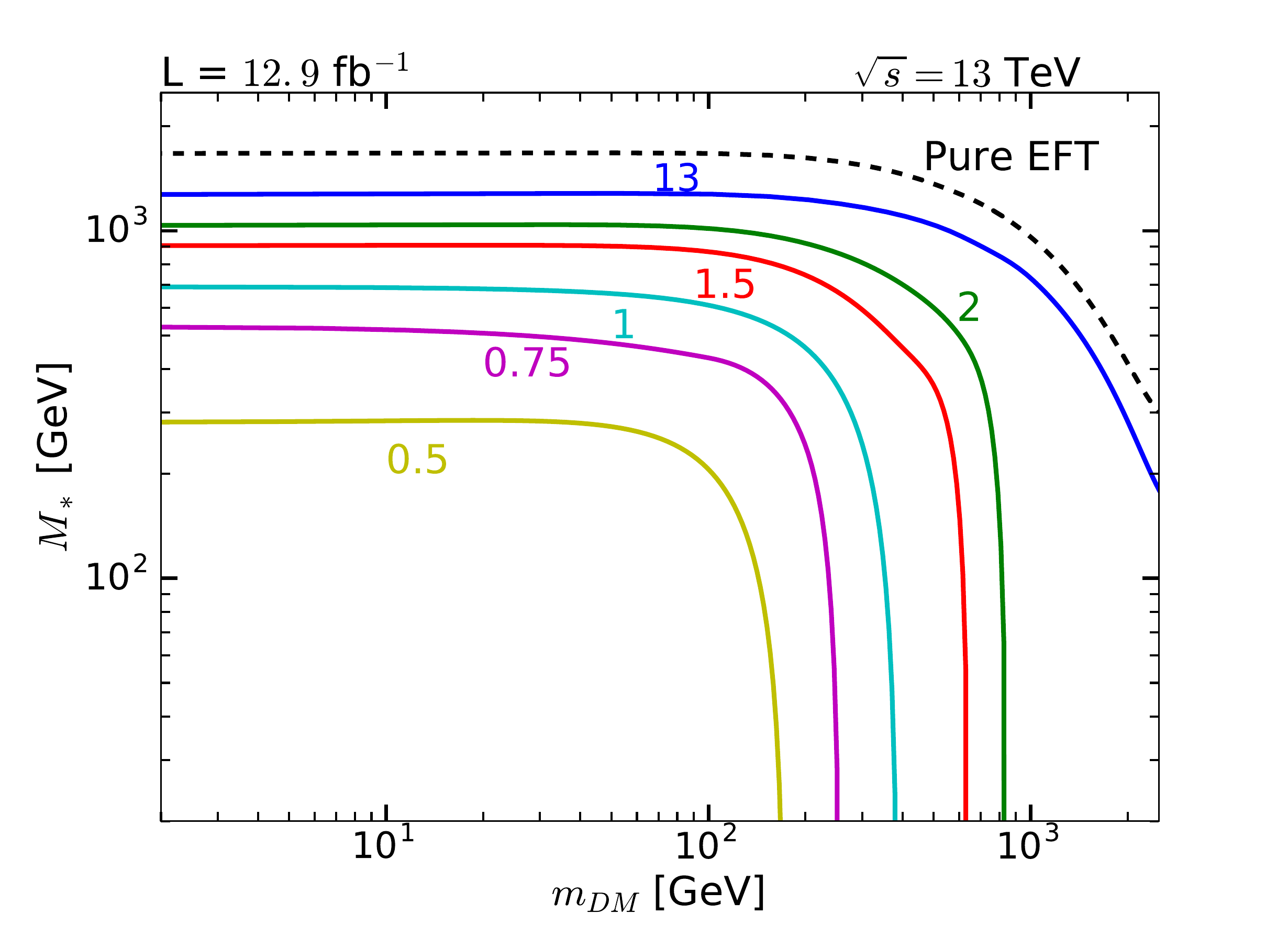}
	\end{minipage}
	\begin{minipage}[b]{8cm}
	\centering
	\includegraphics[width=8cm]{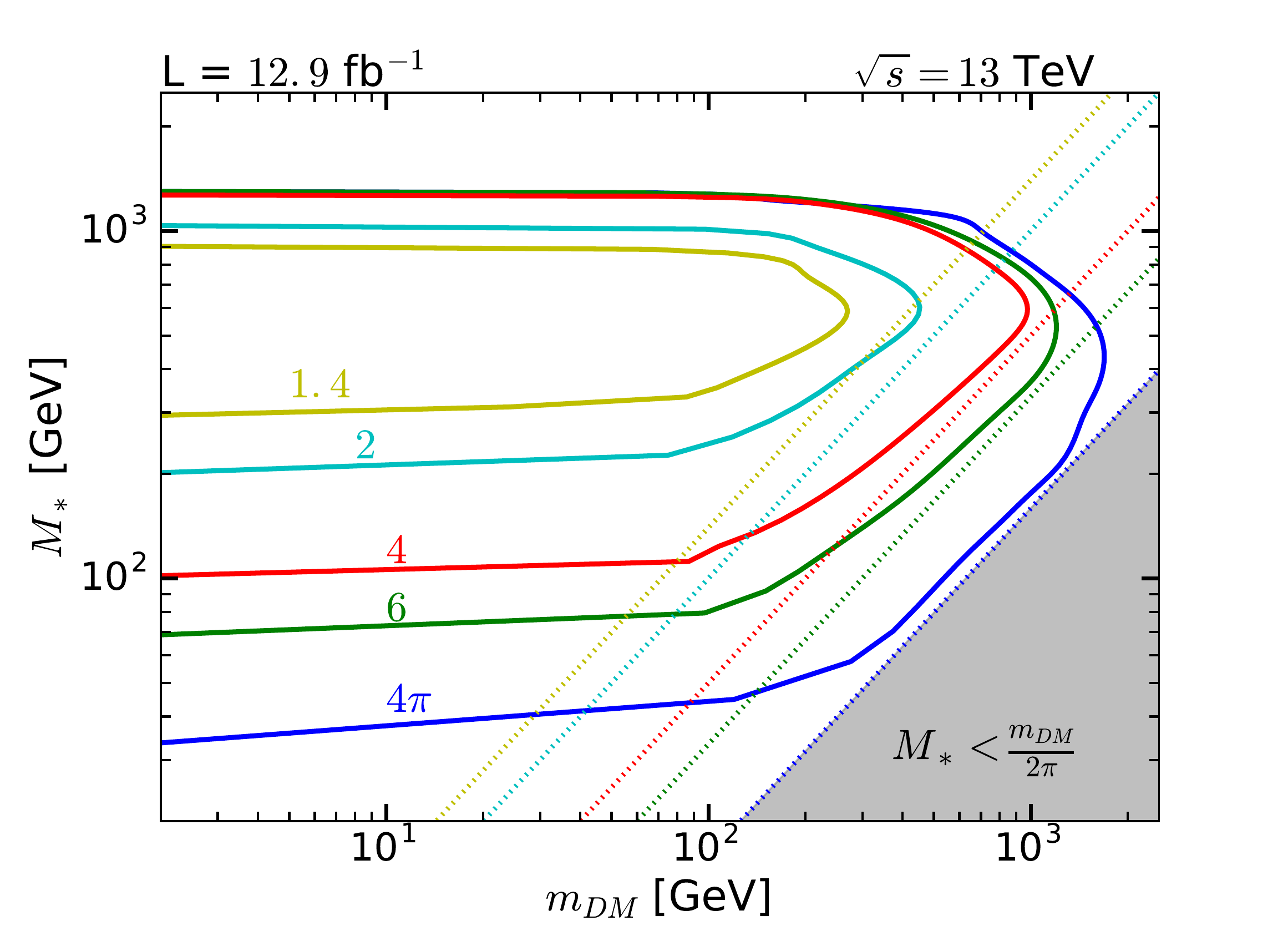}
	\end{minipage}
	\caption{Limits in the $\mdm$-$M_*$ plane at fixed $M_{\textrm{cut}}$ (left panel) and $g_*$ (right panel), obtained from the reinterpretation of ref.~\cite{Sirunyan:2017hci}. The dashed lines on the right plot represent the boundaries of the theoretically forbidden region $2\,\mdm>M_{\textrm{cut}}=g_*M_*$ at fixed $g_*$. In this region, DM production never occurs within the range of validity of the EFT and no limit can be set. The meaning of the ``Pure EFT'' line on the left plot is explained in the text.}  \label {limits}
\end{figure}

All what is missing to compute the $\chi^2$ in eq.~(\ref{chisqt}) are the signal central values $\S_i^0$, which we obtain by simulations and parametrise by the obvious generalisation of eq.~(\ref{crossxEFT})
\begin{equation}\displaystyle
\label{crossxEFTbins}
\S_{i}^0 (M_*, \mdm, M_{\textrm{cut}}) = {\mathcal{L}}\cdot \left [ \frac{1\,{\textrm{TeV}}}{M_*} \right ]^4 \cdot \overline{\sigma} (\mdm) \cdot \epsilon_i (\mdm,  M_{\textrm{cut}})\,, 
\end{equation}
where $ {\mathcal{L}}=12.9$~fb$^{-1}$ is the nominal integrated luminosity and $\epsilon_i$ is the efficiency in each bin. Efficiencies are calculated as interpolating functions in the $\mdm$-$M_{\textrm{cut}}$ plane, using the $7\times7$ grid of simulations described above. This allows a fast exploration of the parameter space.  

The last step is the minimisation of the $\chi^2$ over the unknown additional signal components $\Delta_i\geq0$. Out of $\chi^2_{\textrm{min}}=t_{\rm{min}}$ one obtains the $\alpha=95\%$ limit by solving eq.~(\ref{pexc}). What plays the role of the signal strength $\mu$ here is $1/M_*^4$, therefore the upper exclusion limits on $\mu$ will be reported as lower limits, $M_*^{\textrm{exc}}$, on the effective interaction scale $M_*$. The minimisation of $\chi^2$ is performed numerically, using as staring point of the minimisation algorithm the minimum of the modified $\chi^2$ in eq.~(\ref{chisqtmod}). Being a quadratic function, the location of the minimum of $\chi^2_{\textrm{mod}}$ can be obtained analytically and it is found to be a very good approximation of the true one. The result of this procedure is shown in figure~\ref{limits} in the $\mdm$-$M_*$ plane at fixed $M_{\textrm{cut}}$ (left panel) and $g_*$ (right panel). On the left plot, all the points below the curves are excluded, while in the one on the right what is excluded is the interior of the curves. This peculiar behaviour \cite{Racco:2015dxa} is due to the fact that at fixed $g_*$ (see eq.~(\ref{gstar})) low values of $M_*$ correspond to small $M_{\textrm{cut}}$ and the analysis is not sensitive to very small $M_{\textrm{cut}}$ because of the $\MET>200$~GeV cut (see eq.~(\ref{thr})) on the signal region. 

The meaning of the black curve on the right panel of figure~\ref{limits}, labeled as ``Pure EFT'' needs to be explained. The statistical treatment of the unknown additional signal, which eventually led us to our limit-setting strategy, aims at taking rigorously into account the possible contamination of the EFT signal due to high-energy DM production, occurring and $E_{\textrm{cm}}>M_{\textrm{cut}}$. However it legitimate to assume that high-energy contributions are small or absent, for instance because the mass of the mediator particles, and in turn the EFT cutoff, is above the total LHC energy of $13$~TeV.\footnote{Actually it is legitimate to assume $M_{\textrm{cut}}>13$~TeV only if $M_*\gtrsim 13/4\pi=1$~TeV, since $M_{\textrm{cut}}\lesssim 4\pi M_*$ by perturbativity. Given that $M_*=1$~TeV is right at the boundary of the region that can be excluded by the analysis, $M_{\textrm{cut}}>13$~TeV is a rather marginal configuration. However the considerations that follow also apply to mediators lighter than $13$~TeV but still in the multi-TeV range such that they are too heavy to be produced.} Under this restrictive assumption, no unknown additional signal is present and the ordinary limit-setting strategy can be carried on. The result, obtained with the ``${\widetilde{q}}_\mu$'' test statistic of ref.~\cite{Cowan:2010js} and with the simplified likelihood, is dubbed ``Pure EFT'' limit in figure~\ref{limits}, since it assumes no contamination in the signal region from non-EFT reactions. The Pure EFT limit is stronger than the one obtained with $M_{\textrm{cut}}=13$~TeV using our procedure. This was expected because the ${\widetilde{q}}_\mu$ test is signal-strength based while ours it is not, for the reasons explained in section~\ref{whynot}. Physically this is due to the fact that the Pure EFT test employs the full theoretical information about the shape of the signal distribution, which is partially lost in our analysis because of the additional signal. The difference between the Pure EFT and the $13$~TeV lines in figure~\ref{limits} gives a measure of this effect. Notice that the Pure EFT limit is in some sense more correct than the $M_{\textrm{cut}}=13$~TeV one, because assuming $M_{\textrm{cut}}=13$~TeV guarantees that no additional signal is present. The limit on the $M_{\textrm{cut}}=13$~TeV configuration should thus be set with a signal-strength-based approach that has a stronger expected limit. For lower $M_{\textrm{cut}}$ instead additional signal must be taken into account and our limit-setting strategy is the only viable one.

\subsection*{Validation}

\begin{figure}
\centering
\includegraphics[width= 8cm]{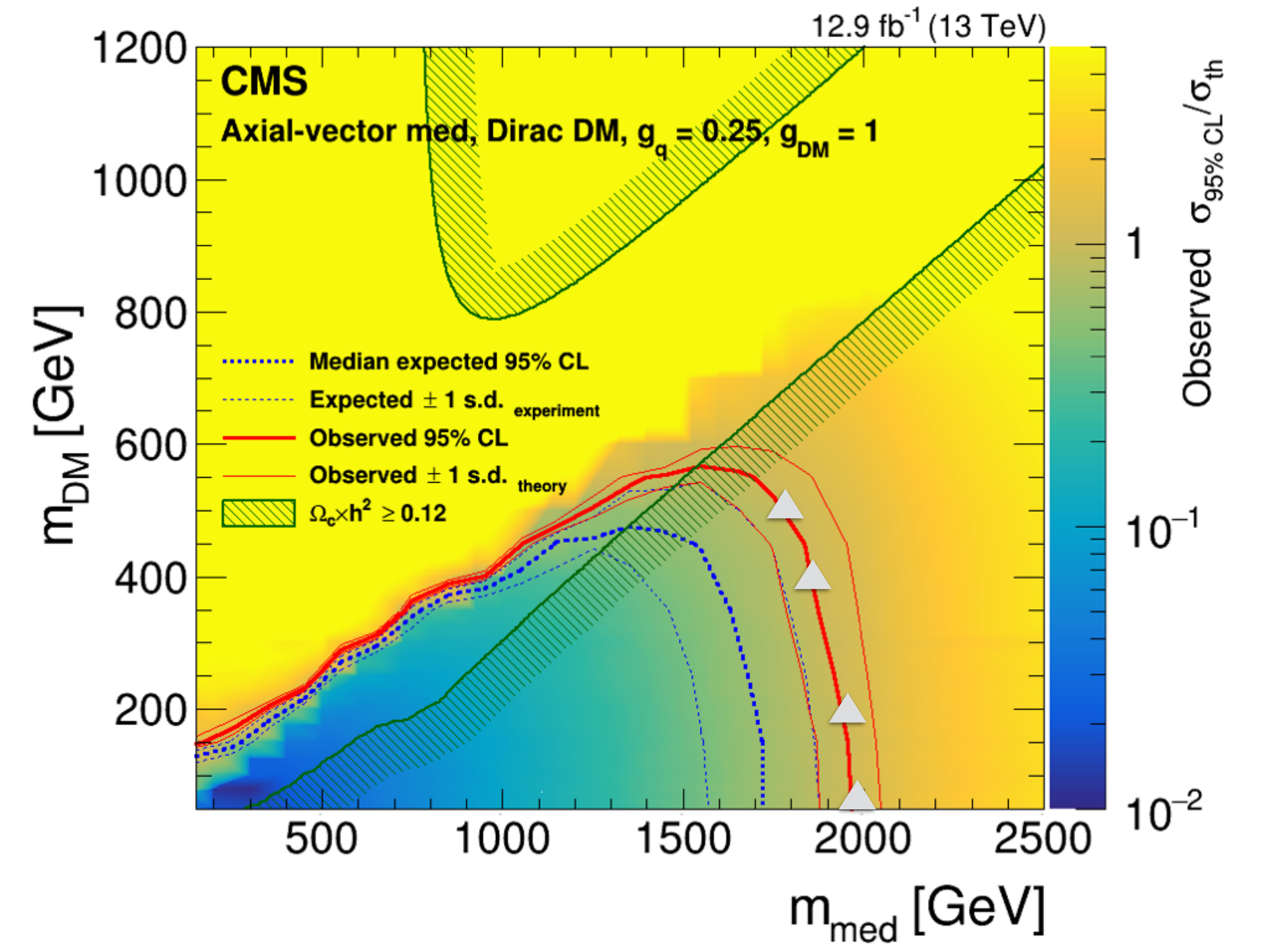}
\caption{The grey triangles, superimposed on the CMS limit, show $95\%$ C.L. lower limits on the mediator mass $m_{\textrm{med}}$ for some values of $\mdm$. The figure is obtained within the benchmark model employed by CMS in ref.~\cite{Sirunyan:2017hci} and for benchmark values, reported on the plot, of its parameters.} 
\label{simplified_confronto}
\end{figure}

Our results rely on two main approximations: the Simplified Likelihood approach and the usage of {\sc{Delphes}} for detector simulation. Both these tools have been validated in the literature, however it is worth cross-checking. We do so by reproducing, using the same tools, the limits on a conventional DM benchmark model (see \cite{Sirunyan:2017hci} for details) where DM is a Dirac fermion coupled to a spin-$1$ $Z^\prime$-like mediator. The results, obtained with the ``${\widetilde{q}}_\mu$'' test statistic \cite{Cowan:2010js}, are reported as grey triangles in figure~\ref{simplified_confronto}, showing perfect agreement. 

\section{Conclusions}\label{conc}

The absence so far of direct discoveries makes the indirect exploration of heavy new physics by means of EFT's one of the priorities of the LHC experimental program. Significant progresses in this direction are expected from run-$2$ and run-$3$ data, and even more from the High-Luminosity LHC upgrade. It is thus important to clarify all the aspects related with the usage of EFT's in the LHC environment. In this paper we addressed the issue related with the limited range of validity of EFT's and we found a rigorous and simple procedure to set limits on the EFT parameter space, duly enlarged to include the EFT cutoff parameter. A full-fledged treatment, to be carried on by the experimental collaborations, is possible and not more complicated than ordinary hypothesis tests. A Simplified Likelihood procedure can also be constructed, relying on approximations that we analysed in great detail.

The procedure has been applied to the CMS mono-jet search \cite{Sirunyan:2017hci}, which we reinterpreted as limits on axial-axial DM EFT. The resulting plots, shown in figure~\ref{limits}, can be employed in two ways. First, they provide a semi-quantitative assessment of how heavy the mediator particles (i.e., how large $M_{\textrm{cut}}$) need to be for the ``Pure EFT'' limit to hold to good approximation. For instance we conclude that the Pure EFT constraint approximately applies for mediator masses above around $2$~TeV. Second, the plot can be quantitatively reinterpreted in specific UV DM models, where the appropriate value of $M_{\textrm{cut}}$ can be worked out. Concrete examples of this reinterpretation are given in ref.~\cite{Racco:2015dxa}. Notice that the bounds one obtains in this way might not be the strongest experimental limits on the UV model at hand. Especially at low $M_{\textrm{cut}}$, i.e., low mediator mass, mediators are likely to be efficiently constrained by direct searches. UV models will be constrained by combining patches, the EFT limit in figure~\ref{limits} being one of those patches. The right panel of figure~\ref{limits} gives another important semi-quantitative information. Namely that WIMP-like DM models, where $g_*\simeq1$, are poorly constrained by mono-jet DM searches. Direct mediator searches are likely to be more effective in that regime.

The limit-setting strategy we developed can be straightforwardly applied to EFT's aimed at describing the effects of heavy new physics in the EW-plus-Higgs sector. However whether our approach is really needed or not is question that needs to be addressed case-by-case. For a concreteness, consider the EFT study proposed by one of us in ref.~\cite{Farina:2016rws}, namely the search, in neutral ($l^+l^-$) and charged ($l\nu$) Drell-Yan at high mass, of two $d=6$ operators that induce ``oblique'' corrections to the SM vector bosons propagators. For the neutral Drell-Yan analysis there is obviously no need for our statistical procedure because the center of mass energy of the events can be measured experimentally. When this is the case, the limit on the EFT can be set by using only events with $E_{{\rm{cm}}}<{M_{\rm{cut}}}$, employing the standard statistical tools, as explained in the Introduction. This is not the case for the charged Drell-Yan analysis, where the relevant distribution is the one in transverse mass, which can receive contributions from arbitrarily high $E_{{\rm{cm}}}$. In principle, charged Drell-Yan should be treated with our strategy, in full analogy with the DM mono-$X$ searches. There is however a substantial quantitative difference. The charged Drell-Yan analysis has such a powerful reach on the EFT Wilson coefficients that it is conceivable to assume that the mediator particles are as heavy as $10$~TeV (in ``plausible'' Universal models, they could be even much higher if other microscopic origins were considered for the EFT operators). It is thus expected (and could be checked) that the mediator particles cannot be produced at a sufficient rate to contribute significantly to the signal region. The ``Pure EFT'' limit-setting procedure, in which one assumes vanishing additional signal, is thus justified in this case. On the other hand, one might want to study what happens if the mediators are light (but still heavy enough not to be directly seen). This would indeed require applying our limit-setting strategy.

Having in mind limit-setting, in this paper we completely ignored the possibility that a significant excess is observed with respect to the SM prediction. If this occurs, our procedure will obviously return a much larger $p$-value than expected, and the result might be used to exclude the SM by computing the probability of such a high $p$-value in the SM hypothesis. However, differently from what happens for a ``normal'' signal, the dependence of the $p$-value on the signal-strength $\mu$ will not tell us much about its true value. The behaviour of the $p$-value in the case of a discovery is easily understood by looking at eq.~(\ref{tmin}) and assuming that the observed countings are those predicted by the EFT with signal-strength ${\overline{\mu}}\neq0$, plus a certain amount of positive additional signal ${\overline{\Delta}}_i$ in each bin. If the hypothesis $\mu=0$ is considered in eq.~(\ref{tmin}), and if the observed are sufficiently above the SM backgrounds $\B_i$, a {\emph{small}} value of $t_{\textrm{min}}$ will be found, and thus a large $p$-value, because most of the bins are over-fluctuating and do not contribute to the signal. The $\mu=0$ hypothesis would thus be found to be fully consistent with the data, which is correct because the over-fluctuation could very well be due to the additional signal rather than to the EFT contribution. The $p$-value will stay constant and large as $\mu$ increases, until $\mu={\overline{\mu}}$. At that point the minimisation over $\Delta_i$ will of course return $\Delta_i= {\overline{\Delta}}_i$ and perfect compatibility will be found. The $p$-value will start decreasing for $\mu>{\overline{\mu}}$, with a slope that depends on the $O_i$'s, and thus in turn on true values of the additional signal components ${\overline{\Delta}}_i$. It is only when $\mu$ is large enough for $\S_i+\B_i$ to overcome $O_i$ in several bins that large $t_{\textrm{min}}$, and thus small $p$-value, will be obtain obtained. Unlike a ``normal'' $p$-value, ours does not show a peak at $\mu={\overline{\mu}}$, but rather an (not necessarily sharp) end point. It can thus be used to set an upper limit on $\mu$, not to measure its value. Therefore, our statistical analysis is not suited to characterise a discovery in terms of the measurement of the EFT Wilson coefficients. Still the EFT can be used to characterise the discovery under the ``Pure EFT'' assumption $\Delta_i=0$. If this assumption is not self-consistent, or if it is not consistent with data, explicit UV models should be employed to characterise the discovered signal.

\section*{Acknowledgments}

We thank P.~Checchia and D.~Racco for useful discussions. A.~W. acknowledges the MIUR-FIRB grant RBFR12H1MW and the PRAT grant ``Fundamental Physics at the TeV scale''.

\appendix

\section{Toy example}\label{toy}

In order to illustrate the validity of the limit-setting strategies developed in sections~\ref{thp}, \ref{full} and \ref{fs}, we consider an histogram with $N=3$ equally spaced bins, constructed for a variable $x\in[0,3]$. The signal contribution to the expected countings, i.e. the signal from the EFT, is taken to be exponentially distributed with total cross-section $\mu$. The total SM background is called $\B$ and the background distribution is also an exponential, but with slightly different slope. Namely
\beq\displaystyle\label{disttoy}
\frac{d\S}{dx}=\mu\frac{e^{-\lambda_sx}}{1-e^{-3\lambda_s}}\,,\;\;\;\;\;\frac{d\B}{dx}=\B\frac{e^{-\lambda_bx}}{1-e^{-3\lambda_b}}\,,
\eeq
with $\lambda_s=1/5$ and $\lambda_b=2/5$. The slopes are chosen in such a way that the signal-over-background ratio is comparable in the three bins. This avoids the limit being dominated by a single bin, in which case our limit-setting strategy would effectively reduce to the one in eq.~(\ref{singlebin}). We momentarily assume that the signal and the background are exactly known. We will study the impact of nuisance parameters later.

\begin{figure}[t]
\begin{center}
\includegraphics[width=220pt]{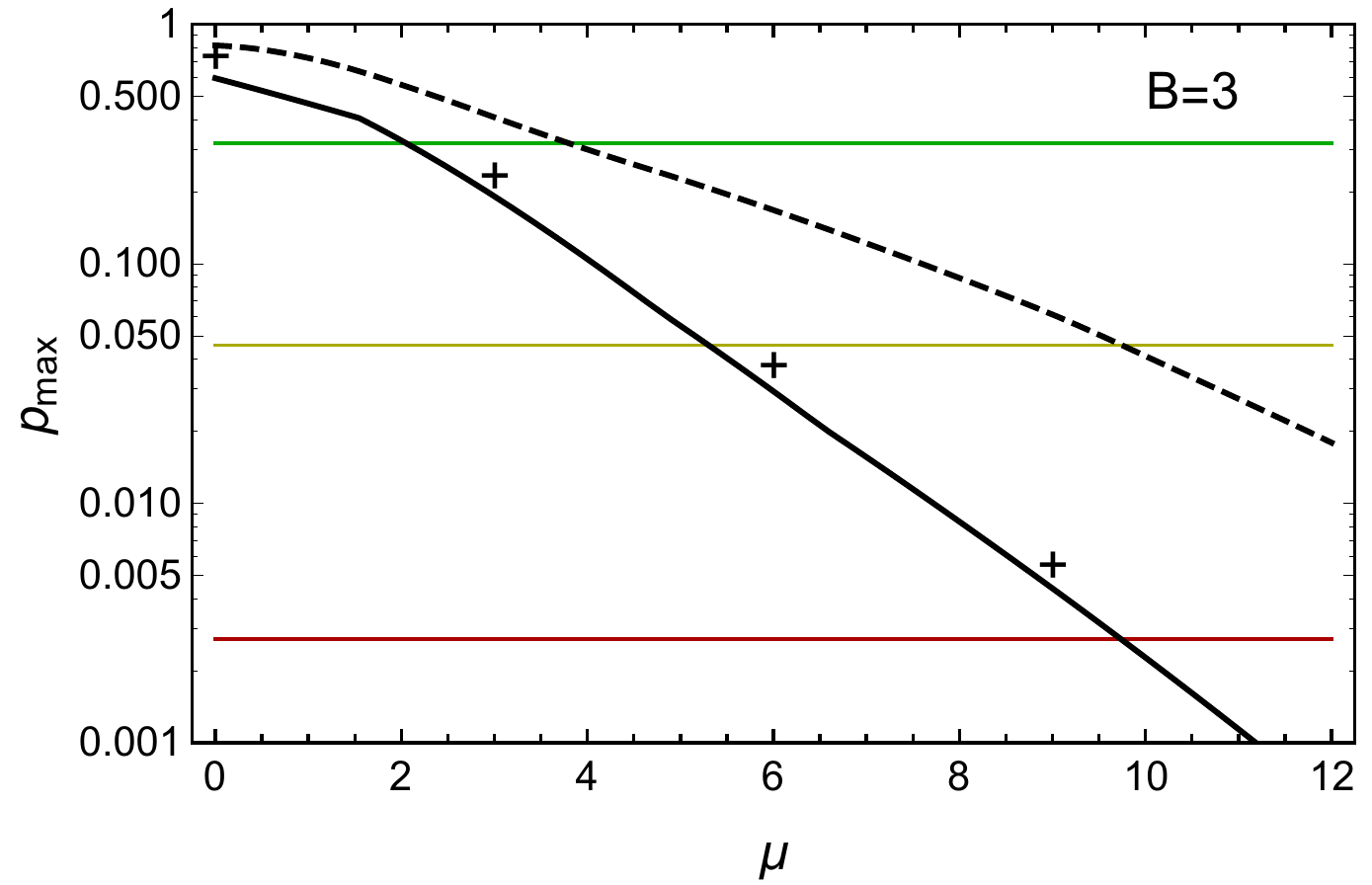}\hspace{10pt}
\includegraphics[width=220pt]{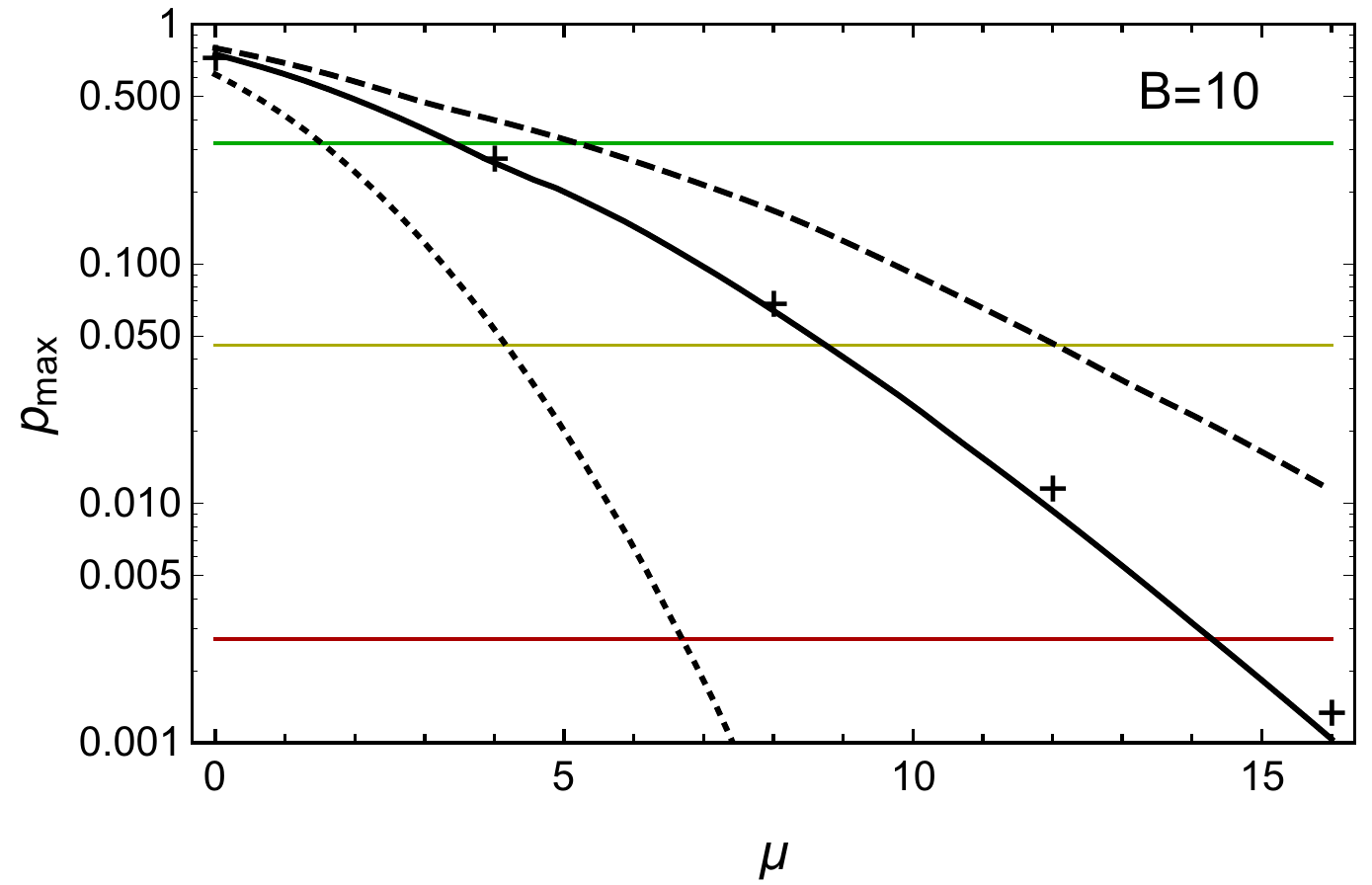}
\caption{The expected $p$-value, computed at the median over $100$ toy Monte Carlos, for $\B=3$ (left) and $\B=10$ (right). The crosses represent the ideal $p$-value, continuos, dashed and dotted lines represent the full, the $\chi^2$ and the modified $\chi^2$ tests, respectively. The levels of $p$-value corresponding to $1$, $2$ and $3\,\sigma$  are also displayed. 
}\label{f1}
\end{center}
\end{figure}

We consider background values $\B=3$ and $\B=10$, which by integrating eq.~(\ref{disttoy}) in the bins lead to $\vec\B\simeq\{1.4,0.95,0.64\}$ and $\vec\B\simeq\{4.7, 3.2, 2.1\}$, respectively. In each case we generated $100$ toy Monte Carlos, i.e. a set of observed $\vec{O}$ vectors distributed according to the background-only hypothesis, and studied the performances of our tests on each of them. The ``ideal'' test defined by eq.~(\ref{pvid}) is obviously the most difficult one to implement, because it requires the determination of the distribution of the test statistic point-by-point in the $\D$-space. Rather than by the Monte Carlo method, we found more convenient to compute the $p$-value directly by summing the probabilities of the possible outcomes of the experiment that are more incompatible than the actual observation $\vec{O}$. Namely, we compute, for each $\mu$, $\D$ and $\o$
\beq\displaystyle
p(\mu,\D;\o\,)=\sum\limits_{o_{1,2,3}=1}^\infty\Theta[t(\mu,\D;\vec{o})-t(\mu,\D;\vec{O})]\prod\limits_{i=1}^3\frac{(\mu\,\overline\S_i+\B_i+\Delta_i)^{o_i}}{o_i!}e^{-{(\mu\,\overline\S_i+\B_i+\Delta_i)}}\,,
\eeq
where $\Theta$ is the step function and the test statistic $t$ is the standard Log Likelihood Ratio, i.e. the argument of the curly bracket in eq.~(\ref{PLR2}). The normalised signal $\overline\S_i\simeq\{0.40, 0.33, 0.27\}$ is defined in eq.~(\ref{truesig}) and is obtained by integrating eq.~(\ref{disttoy}) in the bins. The sum over $o_i$ is of course truncated at a maximal value in the calculation, and this maximum is increased in suitably designed step through an iterative procedure. The procedure stops when the relative change of the $p$-value with respect to the previous step is below $10^{-3}$. Afterwards, $p_{\textrm{max}}(\mu;\o\,)$ needs to be computed by maximising $p$ over $\Delta_i\geq0$. This step is greatly facilitated by taking as starting point of the algorithm the maximum of $t$, which is provided by eq.~(\ref{tmin}). The result of this procedure is reported in figure~\ref{f1} as a function of $\mu$, together with the ``full'' $p$-value (continuous line) defined in section~\ref{full}, the one obtained by the $\chi^2$ formula in eq.~(\ref{chisqt}) (dashed line) and the ``modified'' $\chi^2$ in eq.~(\ref{chisqtmod}) (dotted line). The full $p$-value and the $\chi^2$ are obtained as explained in the main text and don't pose any computational issue. Notice in particular that the full $p$-value can be computed analytically in the present example, where no nuisance parameters are included, thanks again to eq.~(\ref{tmin}). The modified $\chi^2$ formula (\ref{chisqtmod}) is not applicable for $\B=3$, because the expected countings are so low that the observed ones are very likely to have one vanishing entry.

\begin{figure}[t]
\begin{center}
\includegraphics[width=220pt]{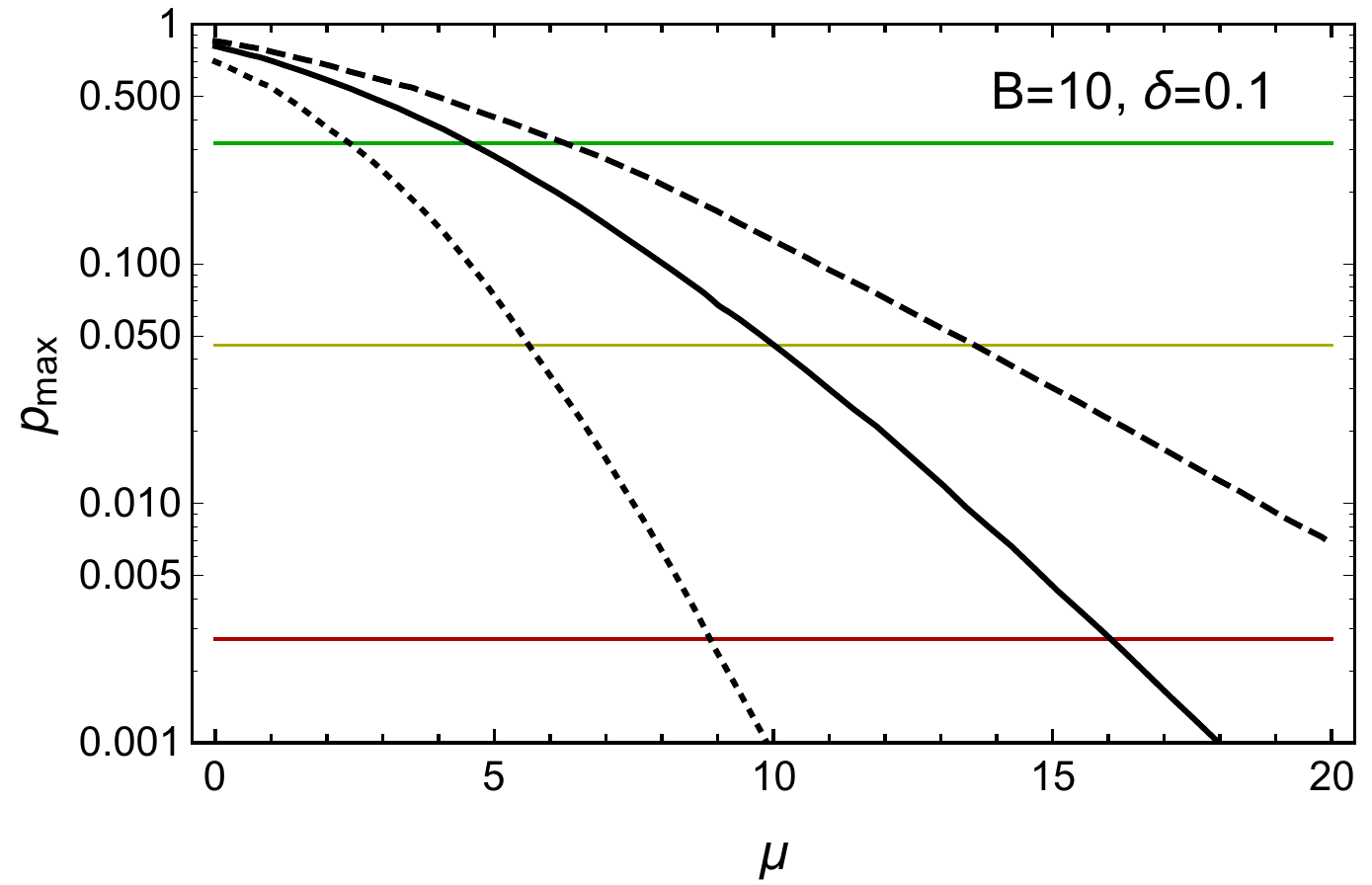}\hspace{10pt}
\includegraphics[width=220pt]{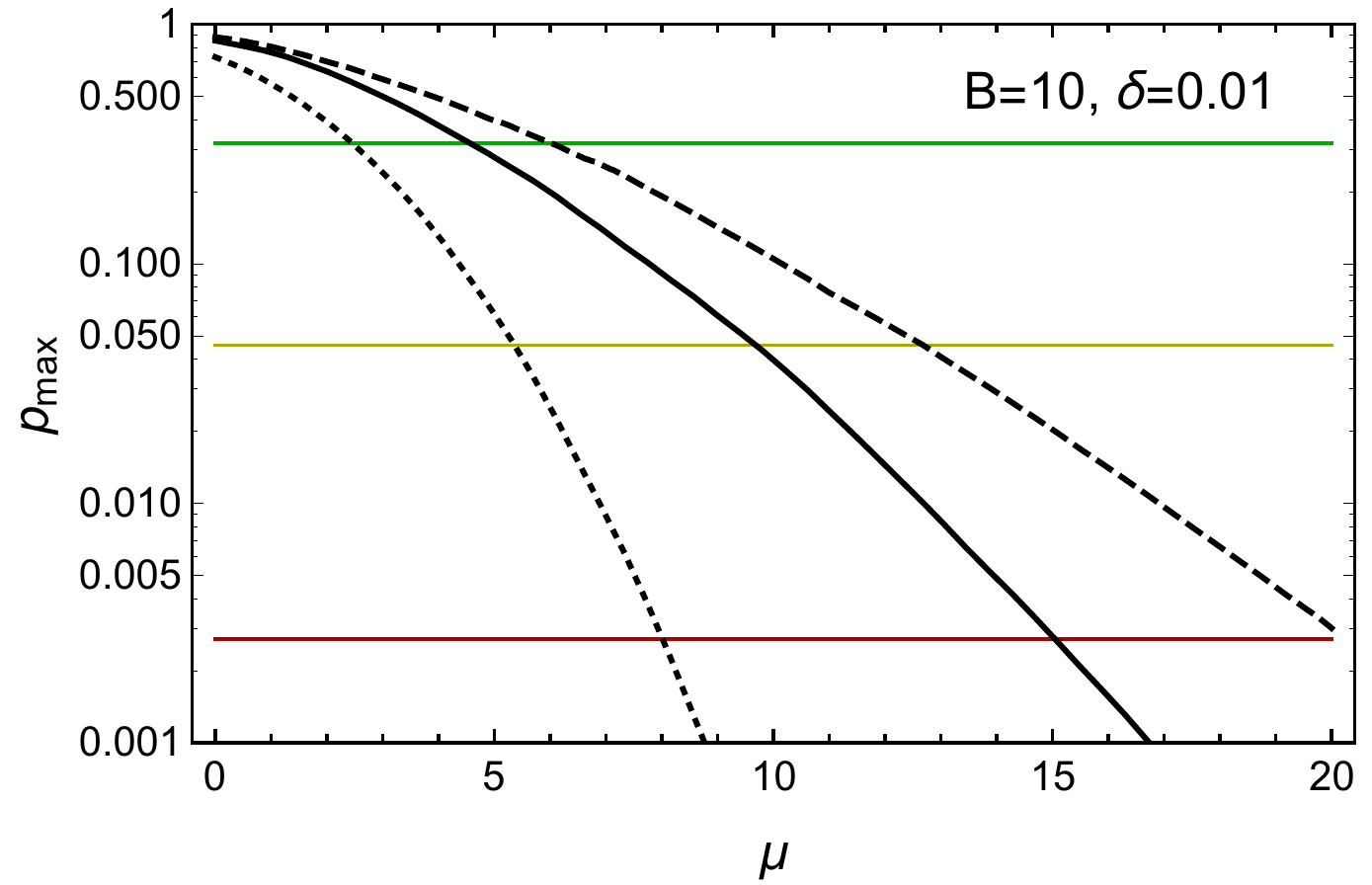}
\caption{The expected $p$-value, computed at the median over $1000$ toy Monte Carlos, for $\B=10$, $\delta=0.1$ (left) and $\delta=0.01$ (right). Continuos, dashed and dotted lines represent the full, the $\chi^2$ and the modified $\chi^2$ tests, respectively. The levels of $p$-value corresponding to $1$, $2$ and $3\,\sigma$  are also displayed. 
}\label{f2}
\end{center}
\end{figure}

Figure~\ref{f1} shows the expected $p$-value, obtained as the median over the possible outcomes, as a function of $\mu$. The agreement of the full version of the test with the ideal one is remarkable, in spite of the fact that the expected background countings are of order one for $\B=3$, while the full test is supposed to hold in the AL where the number of countings is large. This confirms the common lore according to which the statistical distribution of the Log Likelihood Ratio is well-described by the AL formula in eq.~(\ref{chisqN}) even if the number of counting is not large. Other AL formulas are much less accurate. For instance the Gaussian approximation for the Poisson distribution is known to become accurate for a number of countings well above $5$. Indeed we see in the figure that the $\chi^2$ approximation of the $p$-value is completely inaccurate for $\B=3$ and that it is not yet satisfactory (though the accuracy improves) for $\B=10$. The modified $\chi^2$ is found to be completely off, compatibly once again with generic expectations. Notice that the agreement of the expected $p$-value is not strictly speaking sufficient to establish the accuracy of our approximations, which we need to hold for each individual possible outcome of the experiment. We inspected the sample of $100$ toy Monte Carlos and verified that for each of them (including those that display large under- and over-fluctuations with respect to the background-only hypothesis) the level of agreement of the different $p$-value calculation is similar to the one we found for the expected one.

\begin{figure}[t]
\begin{center}
\includegraphics[width=220pt]{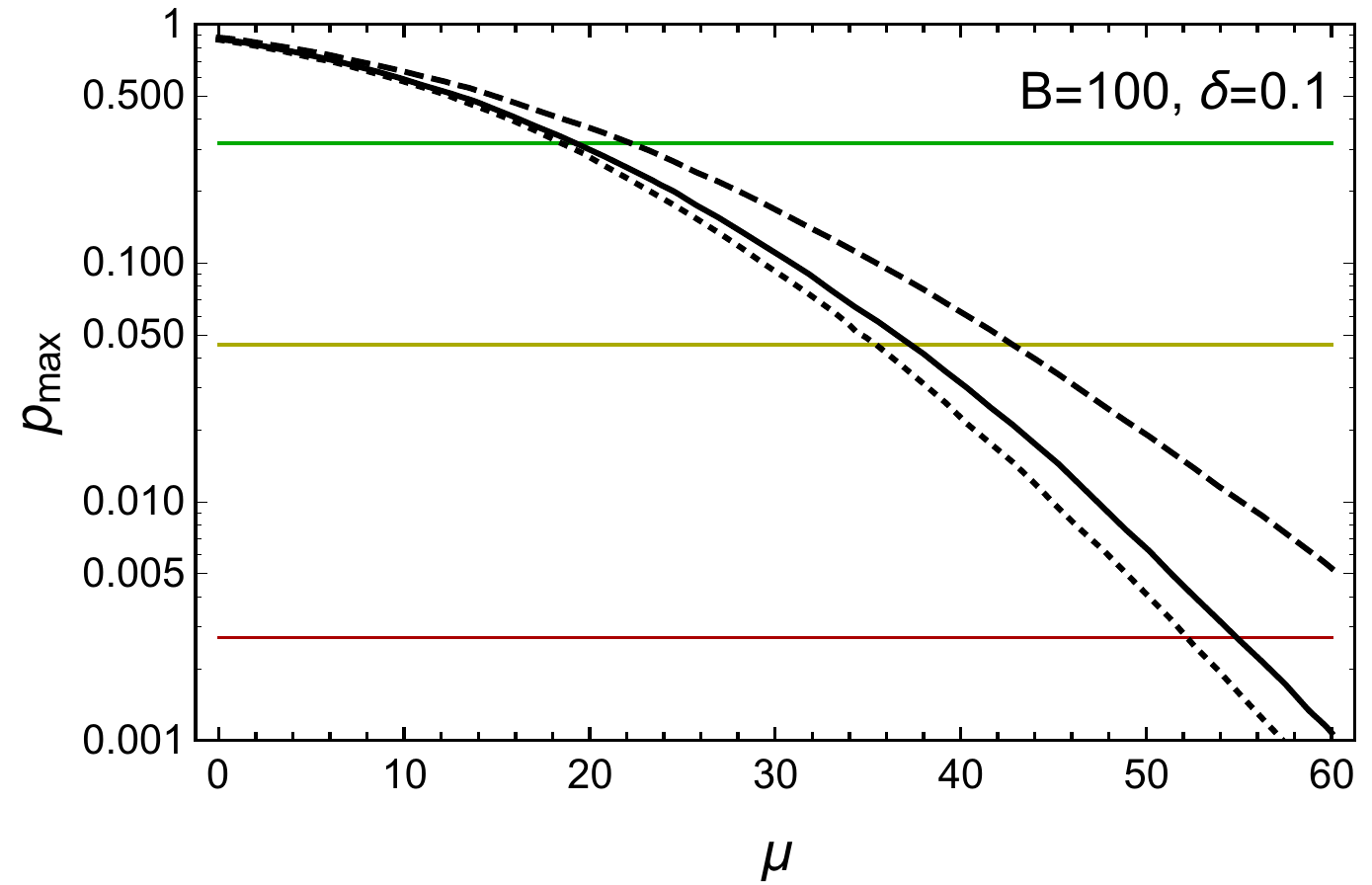}\hspace{10pt}
\includegraphics[width=220pt]{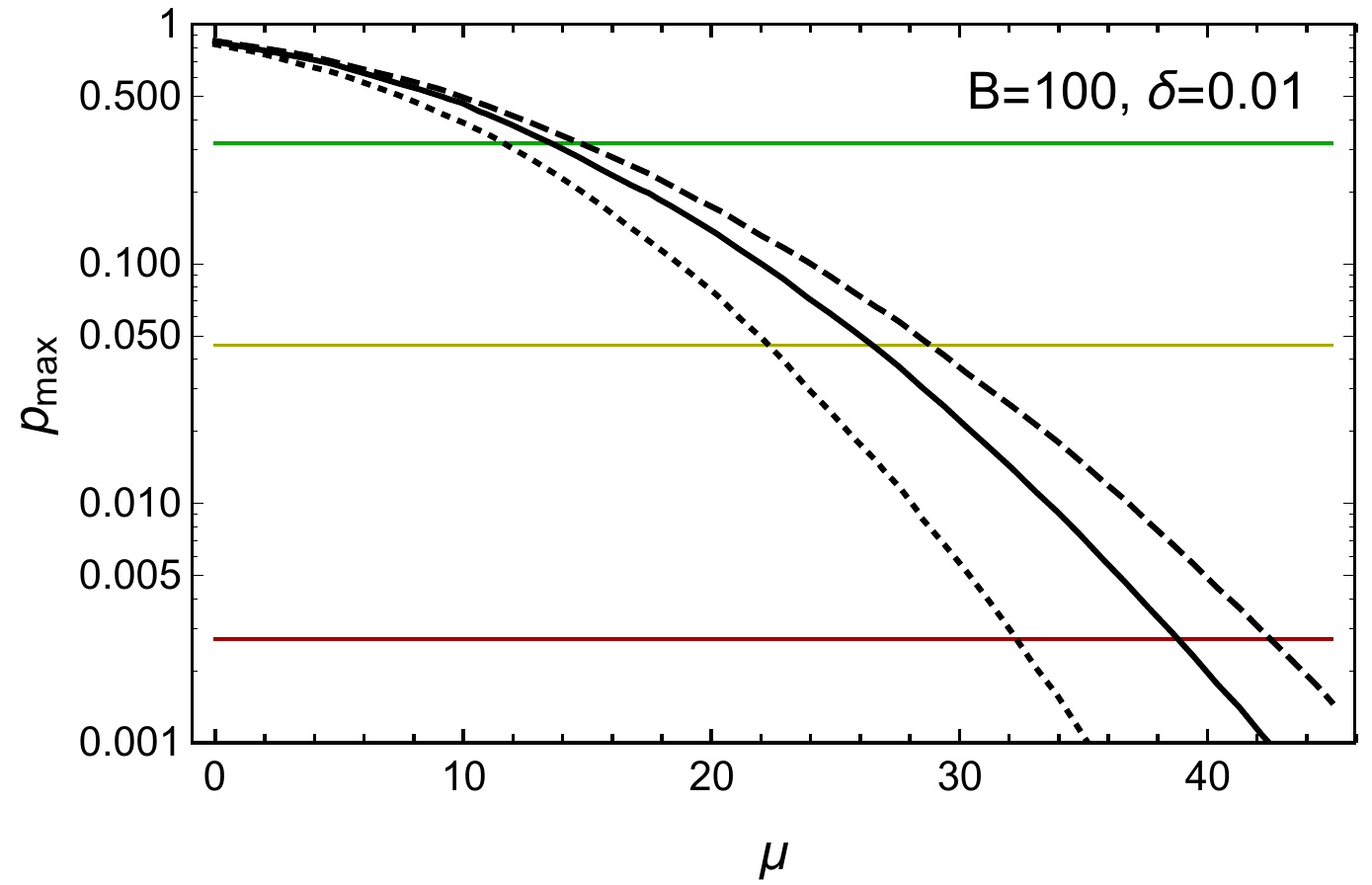}
\caption{The expected $p$-value for $\B=100$, $\delta=0.1$ (left) and $\delta=0.01$ (right).
}\label{f3}
\end{center}
\end{figure}

Next, we complicate our probability model in eq.~(\ref{disttoy}) by adding five sources of nuisance $\n=\{{\hat{\cal{L}}},\beta_1,\beta_2,\beta_3,R\}$, corresponding respectively to the integrated luminosity normalised to the nominal one, to the determination of the background in the $3$ bins from the control region and to the transfer factor needed to relate the latter measurements to the expected background in the signal region. Namely, the signal and background expected countings are taken to be
\beq\displaystyle\label{toynui}
\S_i(\n)=\mu\,{\hat{\cal{L}}}\,{\overline{s}}_i\,,\;\;\;\;\;
\B_i(\n)=R\,\beta_i b_i\,,
\eeq
where ${\overline{s}}_i$ (previously denoted as ${\overline{\S}}_i$) and $b_i$ (previously, $\B_i$) are the signal and background true values obtained by integrating eq.~(\ref{disttoy}) in the bins. The true value of each nuisance parameter is equal to $1$ and its ``measurement'', i.e. the central value of its likelihood, $\nu_i^0$, as obtained by auxiliary measurements, is taken to be Gaussian-distributed with standard deviation $\delta$. In order to mimic the possible outcome of repeated measurements we generated $1000$ points in the $\n^0$ space, with the previously described distribution, and $1000$ toy Monte Carlos for $\o$ distributed around the background true values $b_i$. The covariance matrix $V$ that defines the likelihood of the nuisance parameter in eq.~(\ref{gausnuis}) is $1/\delta^2$ times the identity and it is taken not to fluctuate in the repeated experiments. For each point in the $\n^0$ and $\o$ space we evaluate the full test in eq.~(\ref{tmin}) and the ones based on the $\chi^2$ and the modified $\chi^2$ in eq.s~(\ref{chisqt}) and (\ref{chisqtmod}), as a straightforward application of the formulas in sections~\ref{thp}, \ref{full} and \ref{fs}. The presence of step functions in eq.~(\ref{tmin}) does not pose any computational issue to the automatic minimisation algorithm implemented in {\sc{Mathematica}}. We didn't try to implement the calculation of the ideal $p$-value, which would probably be too demanding already in our toy example. We consider the results obtained in the no-nuisance case sufficient to conclude that the full $p$-value is an accurate estimate of the ideal one, for $\B\geq10$, therefore in what follows we will consider the former as reference to assess the validity of the other methods.

\begin{figure}[t]
\begin{center}
\includegraphics[width=220pt]{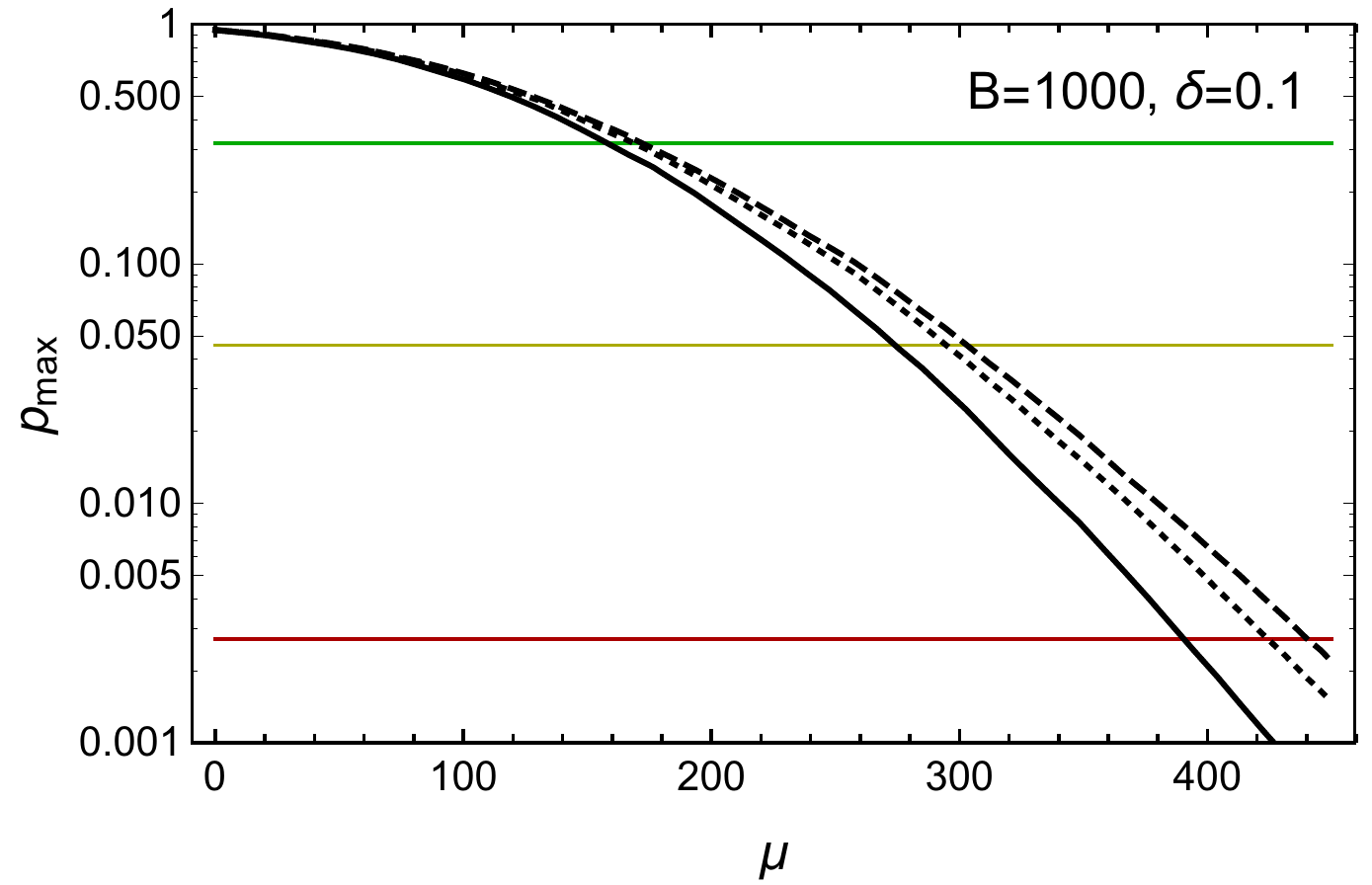}\hspace{10pt}
\includegraphics[width=220pt]{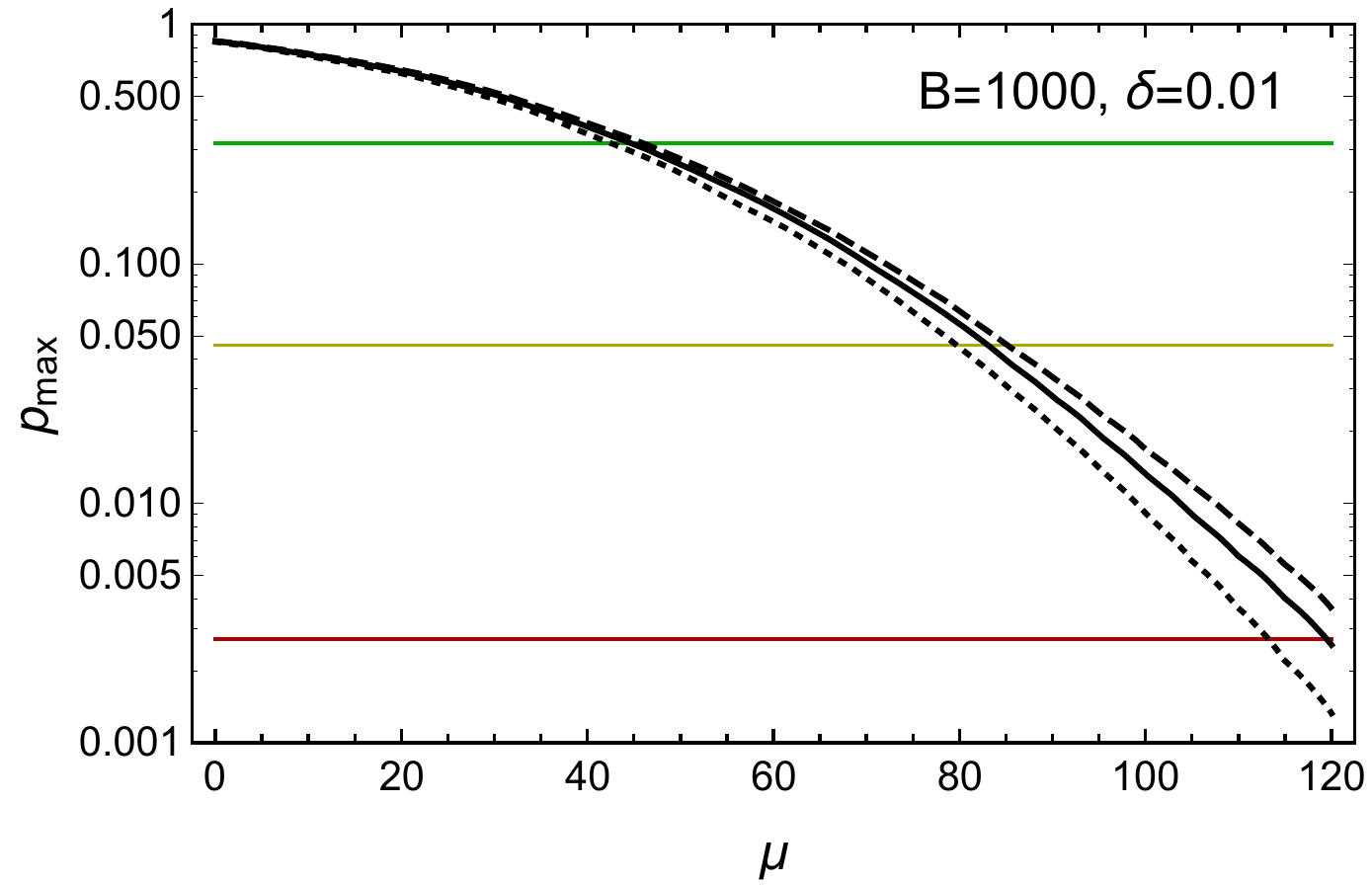}
\caption{The expected $p$-value for $\B=1000$, $\delta=0.1$ (left) and $\delta=0.01$ (right).
}\label{f4}
\end{center}
\end{figure}

Results are obtained for different values of $\B$ and $\delta$, starting from $\B=10$ and $\delta=0.1$ and $\delta=0.01$, shown in figure~\ref{f2}. The level of agreement is comparable to the one in the right panel of figure~\ref{f1} and it does not improve substantially lowering $\delta$. This shows that the un-accuracy of the $\chi^2$ and the modified $\chi^2$ approximations is dominated by the low statistics, while the expansion around $\n=\n^0$ we performed in section~\ref{fs} to obtain the $\chi^2$ formula does not introduce much additional error. The agreement improves for $\B=100$ (see figure~\ref{f3}), because with larger statistics the Gaussian approximation for the Poisson countings becomes more accurate. The un-accuracy of the $\chi^2$ formula due to the $\n=\n^0$ expansion starts becoming visible in this case, resulting in an improvement of the $\chi^2$ approximation (dashed line) when $\delta$ is lowered from $0.1$ to $0.01$. The behaviour of the modified $\chi^2$ $p$-value (dotted line) requires further explanations. For $\delta=0.1$ it provides a rather accurate approximation, but this is an accident due to the fact that for low $\B$ (see $\B=10$, $\delta=0.1$ in figure~\ref{f2}) the modified $\chi^2$ underestimate the true $p$-value while for large $\B$ (see $\B=1000$, $\delta=0.1$ in figure~\ref{f4}) it overestimates it. The point $\B=100$ happens to be close to the transition between the two regimes. Notice also that for $\delta=0.01$ (both at $\B=100$ and $\B=1000$) the performances of the modified $\chi^2$ are inferior to those of the $\chi^2$ as expected. We thus recommend the latter for a more accurate (and conservative, in all cases we studied) estimate of the $p$-value. Finally, the value $\B=1000$ is considered in figure~\ref{f4}. In this case the statistic is large enough for the Gaussian approximation to work extremely well and the expansion around the nuisance central value is the leading source of un-accuracy. The $\chi^2$ approximation becomes indeed substantially exact when $\delta$ is lowered from $0.1$ to $0.01$.

\section{Derivation of the $\chi^2$ formula}\label{appB}
The derivation basically consists in expanding the argument of the ``inf'' eq.~(\ref{PLR2gauss}) up to quadratic order in the nuisance parameters around $\n=\n_0$ as prescribed by eq.~(\ref{Mexp}). Afterwards the minimisation over $\n$ can be performed analytically. The only subtlety is that we expand to the second order only the $(\M_i-O_i)^2$ numerator, while treating the denominator at the zeroth order, i.e. setting $1/\M_i=1/\M_i^0$. This is justified by the fact that the numerator is much more sensitive to the nuisance fluctuations than the denominator. Technically, the contributions to the Taylor series from the expansion of the denominator are proportional to $(\M_i-O_i)^2$, which is small because $\M_i^0-O_i\ll\M_i^0$ for configurations that are not trivially excluded. By proceeding in this way we obtain
\beq\label{t0}
\displaystyle
t=\underset{\n}{\textrm{\large{inf}}}\left\{
c+2\,\vec{b}^{\;t}(\n-\n_0)+(\n-\n_0)^t A (\n-\n_0)
\right\}=c-\vec{b}^{\;t} A^{-1}\vec{b}
\,,
\eeq
where $\vec{b}$ and $A$ are respectively a vector and a matrix in the $\kappa$-dimensional nuisance space, while $c$ is a constant. They read
\bea
&&\displaystyle A = V^{-1}+\sum\limits_{i=1}^N\frac1{\M_i^0}\vec\nabla\M_i^0(\vec\nabla\M_i^0)^t\,,\nonumber\\
&& \displaystyle \vec{b}=\sum\limits_{i=1}^N\frac{\M_i^0-O_i}{\M_i^0}\vec\nabla\M_i^0\,,\nonumber\\
&&\displaystyle c=\sum\limits_{i=1}^N\frac{(\M_i^0-O_i)^2}{\M_i^0}\,.
\eea

In order to evaluate eq.~(\ref{t0}) we need to compute 
\beq\displaystyle\label{t1}
\vec{b}^{\;t} A^{-1}\vec{b}=\sum\limits_{i,j}\frac{\M_i^0-O_i}{\M_i^0}(\vec\nabla\M_i^0)^tA^{-1}\vec\nabla\M_j^0\frac{\M_j^0-O_j}{\M_j^0}\,,
\eeq
which in turn is obtained as follows. Start from the identity
\beq\displaystyle
\Id=A^{-1}A=A^{-1}V^{-1}+\sum\limits_k\frac1{\M_k^0}A^{-1}\vec\nabla\M_k^0 (\vec\nabla\M_k^0)^t\,,
\eeq
and multiply it by $V\vec\nabla\M_j$ on the right and by $(\vec\nabla\M_i^0)^t$ on the left. We obtain a set of scalar equations in the nuisance space, labeled by $i$ and $j$ that run over the $N$ bins. Those can be expressed as a matrix equation in the bin space, namely
\beq\displaystyle\label{mat}
\Sigma_\nu=\Sigma_\M a (\Sigma_\M+\Sigma_\nu)\;\Rightarrow\;\;a=\Sigma_\M^{-1}-(\Sigma_\M+\Sigma_\nu)^{-1}\,,
\eeq
having defined 
\beq\displaystyle
(\Sigma_\M)_{ij}=\M_i^0\delta_{ij}\,,\;\;\;\;\;(\Sigma_\nu)_{ij}=(\vec\nabla\M_i^0)^tV\vec\nabla\M_j^0\,,\;\;\;\;\;(a)_{ij}=\frac{1}{\M_i^0}(\vec\nabla\M_i^0)^tA^{-1}\vec\nabla\M_i^0\frac{1}{\M_j^0}\,.
\eeq
By using eq.~(\ref{mat}) in eq.~(\ref{t1}) we obtain
\beq\displaystyle\label{t2}
\vec{b}^{\;t} A^{-1}\vec{b}=\sum\limits_{i,j}(\M_i^0-O_i)\left[\Sigma_\M^{-1}-(\Sigma_\M+\Sigma_\nu)^{-1}\right]_{ij}(\M_j^0-O_j)\,,
\eeq
from which eq.~(\ref{chisqt}) follows.

\bibliographystyle{mine}
\bibliography{bibliography}

\end{document}